\documentclass[a4paper,12pt]{article}
\pdfoutput=1

\usepackage{jheppub}
\usepackage[utf8]{inputenc}
\usepackage{amsmath,amssymb,amsthm}
\usepackage{physics,mathtools}
\usepackage{tikz,float}
\usepackage{bbm}

\makeatletter
\def\@fpheader{\vspace{0.1mm}}
\makeatother
\newcommand{\bw}{\begin{widetext}}
\newcommand{\ew}{\end{widetext}}
\newcommand{\bea}{\begin{eqnarray}}
\newcommand{\eea}{\end{eqnarray}}
\newcommand{\be}{\begin{equation}}
\newcommand{\ee}{\end{equation}}
\newcommand{\bca}{\begin{cases}}
\newcommand{\eca}{\end{cases}}

\newcommand{\p}{\partial}

\newcommand\Lam{\Lambda}

\newcommand\ra{{\rightarrow}}

\newcommand{\ben}{\begin{enumerate}}
\newcommand{\een}{\end{enumerate}}

\usetikzlibrary{patterns}
\makeatletter
\let\over\@@over
\makeatother

\title{{{Thermal holographic correlators and KMS condition}}}
\author{Ilija Buri\'c,}
\emailAdd{burici@tcd.ie}
\author{Ivan Gusev,}
\emailAdd{gusevi@tcd.ie}
\author{Andrei Parnachev}
\emailAdd{parnachev@maths.tcd.ie}

\affiliation{School of Mathematics and Hamilton Mathematics Institute, Trinity College, Dublin 2, Ireland}

\abstract{
Thermal two-point functions in holographic CFTs receive contributions from two parts. One part comes from the identity, the stress tensor and multi-stress tensors and constitutes the stress-tensor sector.
The other part  consists of  contributions from double-trace operators. The  sum of these two parts must satisfy the KMS condition -- it has to be periodic in Euclidean time. The stress-tensor sector can be computed by analyzing the bulk equations of motions near the AdS boundary and is not periodic by itself. We show that starting from the expression for the stress-tensor sector one can impose the KMS condition to fix the double-trace part, and hence the whole correlator. We perform explicit calculations  in the asymptotic approximation, where the stress-tensor sector can be computed exactly. One can either sum  over the thermal images of the stress-tensor sector and subtract the singularities or solve for the KMS condition directly and perform the Borel resummation of the resulting double-trace data -- the results are the same.
}

\begin{document}

\maketitle

\section{Introduction and Summary}
\label{S:Introduction}

Gravitational holography \cite{Maldacena:1997re,Gubser:1998bc,Witten:1998qj} provides a powerful framework for understanding strongly interacting conformal field theories (CFTs) at large central charge $C_T$, and more generally, quantum field theories. A particularly promising direction is thermal holography, where CFTs at finite temperature $T = 1/\beta$ are mapped to asymptotically anti-de Sitter black holes \cite{Witten:1998zw}. In this context, finite-temperature CFT correlators are related to fields propagating in such spacetimes. A natural question is to investigate the operator product expansion (OPE) of these correlators and their decomposition into finite-temperature conformal blocks. It is also natural to introduce the complexified time $\tau = i t + \tau_R$ and explore questions regarding the analytic behavior of the OPE series in the complex $\tau$-plane.

At leading order in the inverse central charge expansion, finite-temperature holographic two-point functions of identical primary scalar operators
$\phi(x^\mu)$ receive contributions from the identity, the stress tensor and its products (the stress-tensor sector).
In two spacetime dimensions, these contributions constitute the entire correlator. However, in higher dimensions, double-trace operators $[\phi\phi]_{n,\ell} \simeq : \phi\Box^n \p_{\mu_1} \ldots \p_{\mu_\ell} \phi:$ also play a role
(see e.g. \cite{Fitzpatrick:2012yx,Komargodski:2012ek} for the discussion of such operators in the CFT context and \cite{Iliesiu:2018fao,Kulaxizi:2018dxo,Fitzpatrick:2019zqz,Parisini:2023nbd,Ceplak:2024bja} for their appearance in the holographic examples involving asymptotically AdS black holes).

Any finite-temperature correlator satisfies the KMS condition, stating that the correlator is periodic in real $\tau$ with period $\beta$, \cite{Kubo:1957mj,Martin:1959jp}. Additionally, correlators are expected to be analytic in the strip $0<\Re \tau <\beta$. The simplest example of a correlator meeting these conditions is provided by the generalized free field (GFF) model, where the vacuum result is summed over thermal images; see e.g. \cite{Iliesiu:2018fao}. The OPE decomposition of this sum is straightforward: besides the identity's contribution (the vacuum term), the sum over thermal images fixes the double-trace operator contributions. Another perspective is that the double-trace contributions organize themselves precisely to ensure that the correlator satisfies the KMS condition.

The GFF model serves as the simplest toy model of holography, in which the entire stress-tensor sector is replaced by just the identity contribution. As explained above, in this simplified scenario, recovering the double-trace portion of the correlator, which ensures the KMS periodicity, is straightforward. A natural question arises: does this reasoning extend to actual holography? Specifically, given the stress-tensor sector, can one similarly reconstruct the double-trace contributions to the correlator as straightforwardly as in the GFF case?

This question has been implicitly raised in several recent works \cite{Ceplak:2024bja,Parisini:2023nbd,Fitzpatrick:2019zqz}. Technically, controlling the OPE terms due to double-trace operators is significantly more challenging compared to terms arising from the stress-tensor sector. The latter can be effectively obtained by analyzing the bulk equations of motion in the near-boundary expansion, \cite{Fitzpatrick:2019zqz}, with results cross-checked against the momentum space OPE expansion \cite{Karlsson:2022osn,Parisini:2023nbd,Manenti:2019wxs}. However, the method of \cite{Fitzpatrick:2019zqz} does not determine the double-trace terms explicitly, although their existence can be inferred from poles in the OPE coefficients of multi-stress-tensor operators, which appear for integer conformal dimensions of the external scalar fields. These poles signify mixing between multi-stress tensors and double-trace operators. From a momentum-space perspective, double-trace contributions correspond to contact terms, making them difficult to recover directly.

In \cite{Ceplak:2024bja}, the asymptotic behavior of the stress-tensor sector terms in the OPE was analyzed\footnote{See also 
\cite{El-Showk:2011yvt,Katz:2014rla,Li:2019tpf,Fitzpatrick:2020yjb,Alday:2020eua,Rodriguez-Gomez:2021pfh,Rodriguez-Gomez:2021mkk,Karlsson:2021duj,Dodelson:2023vrw,Marchetto:2023xap,David:2023uya,Benjamin:2023qsc,Barrat:2024fwq,Barrat:2024aoa} for additional recent papers where the OPE has been discussed in the context of holographic dual of asymptotically AdS black holes.}. The radius of convergence of the OPE of the stress-tensor sector was shown to be finite, the latter exhibiting singularities in the  strip $0<\Re \tau <\beta$. These singularities, referred to as ‘bouncing singularities' in \cite{Ceplak:2024bja}, correspond to null geodesics bouncing off the singularities of asymptotically AdS black holes in the gravitational description\footnote{See also 
\cite{Kraus:2002iv,Fidkowski:2003nf,Festuccia:2005pi,Engelhardt:2014mea,Horowitz:2023ury,Parisini:2023nbd,DeClerck:2023fax}
for related work.}. Since the complete correlator must not exhibit such singularities, the double-trace terms must display singular behavior precisely canceling bouncing singularities arising from the stress-tensor sector.

The finite radius of convergence ($\beta/[2 \sin (\pi/ d)]$ in $d$ spacetime dimensions) makes it clear that a simple summation over images of the stress-tensor sector is not straightforward, as one needs to analytically continue the stress-tensor sector to get its value at thermal images of $\tau=0$. Moreover, the analysis in \cite{Ceplak:2024bja} only captures the leading singularity of this sector, and the full analytic structure is expected to be considerably more intricate. A natural question arises: how accurately does the asymptotic behavior derived in \cite{Ceplak:2024bja} approximate the entire stress-tensor sector? While this approximation is clearly accurate near the poles (or equivalently, for multi-stress tensors $[T_{\mu\nu}]^n$ with $n\gg 1$), even at the level of the stress tensor itself, where the asymptotic approximation error is greatest, the deviation between the asymptotic behavior and the exact one-point function times the OPE coefficient is relatively small.

This prompts us to introduce the \textit{asymptotic model} of holography, a model where all multi-stress-tensor contributions, beginning with the single stress tensor, are captured by the asymptotic formula. As we will see below, this model can often be explicitly summed, analytically continued, and then summed over thermal images. While the resulting expression is inherently KMS-periodic, it typically exhibits poles (bouncing singularities) within the vertical strip. Nonetheless, it is possible to add a term that affects only the double-trace data, preserves KMS-periodicity, and removes these poles. The resulting  expression  provides a significantly improved approximation to the true holographic correlator compared to the GFF formula. 
Does this procedure fix the double-trace part uniquely?
We believe the answer is yes, under the assumption that we cannot introduce additional double-trace terms which would be non-analytic in the strip\footnote{This is because such an additional double-trace piece is a function of $\tau^2$ and hence is invariant under $\tau\,\ra\,-\tau$. It is therefore periodic under $\tau\, \ra\, \tau+\beta$. Hence, if it's analytic in the strip, it is also analytic in the entire complex plane and therefore is at most a constant. The constant can be set to zero by requiring that the correlator decays at large times.}. 

The asymptotic model just described serves as a good starting point for computing the full holographic correlator, which, we hope, can be achieved by incorporating subleading terms in the stress-tensor sector.

One may wonder if there is an alternative procedure to summing over images. A promising route emerges from analyzing the structure imposed by the KMS condition. In \cite{Marchetto:2023xap} it was pointed out that the KMS condition, which essentially demands vanishing of the odd derivatives at $\tau=\beta/2$, forms an infinite system of linear equations for the thermal one-point coefficients of double-trace operators. However, these equations admit only asymptotic (non-convergent) solutions when the input CFT data is given by the multi-stress-tensor sector. To make sense of these divergent series, we employ Borel regularization.

Remarkably, this regularization yields physically meaningful results. When applied to the GFF, Borel summation reproduces the exact double-trace data. We then apply the same approach to the asymptotic model of holography and find that Borel summation reconstructs the same double-trace data that follows from summing over images and canceling the poles associated with bouncing singularities. Our main results are as follows:

\begin{itemize}
    \item We explicitly construct the holographic correlator in the asymptotic approximation for half-integer values of $\Delta_\phi$,  incorporating the large-$n$ behavior of multi-stress-tensor coefficients. The resulting correlator satisfies the KMS condition, is analytic in the vertical strip, up to the pole at $\tau=0$, and  captures the asymptotic stress-tensor sector structure expected from holography.
    
    \item We review \emph{KMS sum rules}, \cite{Marchetto:2023xap}, an infinite system of linear equations for thermal OPE coefficients arising from the vanishing of odd derivatives of the two-point function at $\tau=\beta/2$. These sum rules generically admit only divergent solutions when the input data comes from the stress-tensor sector.
    
    \item We apply \emph{Borel summation} to regularize the divergent solutions of the sum rules. In the case of the GFF model, this reproduces the previously known expressions for the double-trace coefficients. For the asymptotic  model of holography, the Borel-resummed solution agrees precisely with the result obtained by explicit singularity cancellation, providing a second, independent derivation of the double-trace data.
    
    \item We explore \emph{refinements of the asymptotic model} by incorporating a finite number of exact low-lying multi-stress-tensor coefficients. While these corrections improve the accuracy of the model's input, they must be implemented carefully to preserve analyticity in the complex $\tau$-plane. We find that only a small number of such corrections maintain the desired structure; attempts to go further lead to divergences of the correlator at $\text{Im}\,\tau\to\infty$ unless a more systematic large-$n$ expansion is employed.
\end{itemize}
\smallskip

The paper is organized as follows.
In Section \ref{S:Toy models of holography}, we introduce the asymptotic model of holography, beginning with a general discussion of thermal two-point functions and their conformal block decomposition. We then review the multi-stress-tensor OPE coefficients and proceed to construct the asymptotic model via resummation and the method of images. The section concludes with a discussion on possible refinements of the model by incorporating exact low-lying multi-stress-tensor data. Section
\ref{S:Sum rules} focuses on the derivation and solution of the KMS sum rules. We begin by reviewing the sum rules and deriving the associated system of equations. We then solve these equations, both in the homogeneous case and for particular solutions using Borel summation. This method is applied first to the generalized free field model and subsequently to the asymptotic holographic model, leading to an independent reconstruction of the double-trace data.
In Section \ref{S:Discussion}, we discuss the implications of our results and outline possible directions for future refinements of the asymptotic model.
Appendices \ref{sec:asymptotic-models-delta-d}–\ref{A:Laplace equation on the AdS-Schwarzschild black hole} contain supplementary material: Appendix \ref{sec:asymptotic-models-delta-d} presents asymptotic models for various values of dimension of the external field $\Delta_\phi$ and the spacetime dimension $d$; Appendix \ref{Solutions to Sum Rules} contains details about solutions to the sum rules, including identities used throughout the main text; Appendix \ref{A:Laplace equation on the AdS-Schwarzschild black hole} provides background on the Klein-Gordon equation in the AdS-Schwarzschild black hole geometry, which is central to the computation of the multi-stress-tensor and double-trace coefficients.

\section{Asymptotic model of holography}
\label{S:Toy models of holography}

In this section, we construct the asymptotic model of holography. The first subsection reviews the general properties of conformal two-point functions at finite temperature. In the second subsection, we recall the asymptotic form of the multi-stress-tensor CFT data. The next two subsections are devoted to the computation of the asymptotic model two-point function. The final subsection discusses refinements of the model that take into account more detailed information about the stress-tensor sector.

\subsection{Thermal two-point functions}
\label{SS:Properties of two-point functions}

The main object of study in this work are conformal two-point functions of scalar fields at finite temperature $T = \beta^{-1}$,
\begin{equation}\label{two-point-function}
    g_\beta(\tau,\vec{x}) = \langle \phi(\tau,\vec{x}) \phi(0)\rangle_\beta\ .
\end{equation}
Equivalently, these can be thought of as correlators on the geometry $S^1_\beta \times \mathbb{R}^{d-1}$. The coordinates on $S^1_\beta \times \mathbb{R}^{d-1}$ are denoted by $(\tau,\vec{x})$ and the Euclidean time is periodic, $\tau\sim\tau+\beta$. For the most part, we shall consider the two-point correlation function evaluated at vanishing spatial separation, that is, at $\vec{x} = 0$. This will be denoted by $g_\beta(\tau)$. We begin by listing some of the main properties of \eqref{two-point-function}, starting with those present in any CFT, before turning to the ones specific to theories with holographic duals.

\paragraph{Periodicity and KMS condition} In general, we will allow $\tau$ to take complex values
within the vertical strip $0\leq\Re \tau \leq\beta$.
The KMS condition can be written as
\begin{equation}\label{KMS-condition}
    g_\beta \left(\frac{\beta}{2}+\tau,\,\vec{x}\right) = g_\beta \left(\frac{\beta}{2} - \tau,\,\vec{x}\right)\ .
\end{equation}

\paragraph{Conformal block decomposition} The two-point function can be expanded in conformal blocks as, \cite{Iliesiu:2018fao},
\begin{equation}\label{conformal-block-decomposition}
  g_\beta(\tau,\vec{x}) = \sum_{\mathcal{O}} \frac{\ell_{\mathcal{O}}!}{2^{\ell_{\mathcal{O}}}(\frac{d-2}{2})_{\ell_{\mathcal{O}}}}\frac{\lambda_{\phi\phi\mathcal{O}}b_\mathcal{O}}{\beta^{\Delta_\mathcal{O}}} C_{\ell_\mathcal{O}}^{(\frac{d-2}{2})} \left(\frac{\tau}{|x|}\right) |x|^{\Delta_{\mathcal{O}}-2\Delta_\phi}\,, \quad |x| = \sqrt{\tau^2+|\vec{x}|^2}\ .
\end{equation}
The sum runs over all operators $\mathcal{O}$ that appear in the $\phi\times\phi$ OPE, their dimensions and spins being denoted by $(\Delta_{\mathcal{O}},\ell_{\mathcal{O}})$. The $C^{(\alpha)}_\ell$ are Gegenbauer polynomials, $\lambda_{\phi\phi\mathcal{O}}$ the OPE coefficients of the CFT and $b_{\mathcal{O}}$ the thermal one-point coefficients. 
At zero spatial separation, we may perform the sum over spins $\ell_{\mathcal{O}}$ in \eqref{conformal-block-decomposition} and write
\begin{equation}\label{conformal-block-decomposition-zero-x}
  g_\beta(\tau) = \sum_{\Delta_\mathcal{O}} \frac{a_{\Delta_\mathcal{O}}}{\beta^{\Delta_{\mathcal{O}}}} \tau^{\Delta_{\mathcal{O}}-2\Delta_\phi}\,, \qquad a_\Delta \equiv \sum_{\mathcal{O} \in \phi \times \phi}^{\Delta_\mathcal{O}=\Delta} \frac{\ell_{\mathcal{O}}!\, C^{(\frac{d-2}{2})}_{\ell_\mathcal{O}}(1)}{2^{\ell_{\mathcal{O}}}(\frac{d-2}{2})_{\ell_{\mathcal{O}}}}\lambda_{\phi\phi\mathcal{O}}b_\mathcal{O}\ .
\end{equation}
By a slight abuse of terminology, we shall refer to coefficients $a_{\Delta_\mathcal{O}}$ rather than $b_\mathcal{O}$ as thermal one-point coefficients.

\paragraph{Boundedness} In the complex $\tau$-plane, the two-point function is bounded along the lines parallel to the imaginary axis by its value on the real axis, i.e.
\begin{equation}\label{Regge-bound}
    g_\beta(\tau) \geq |g_\beta(\tau + i \kappa)|\,, \qquad \tau,\kappa\in\mathbb{R}\ .
\end{equation}
We briefly review the derivation of \eqref{Regge-bound} (for similar arguments, see for instance \cite{Fidkowski:2003nf}). Recall that, on the cylinder, the Hamiltonian of the theory coincides with the dilation operator $D$. In Euclidean time $\tau$, the dilation acts as $D = \partial_\tau$, and thus
\begin{equation}
    \phi(\tau) = e^{\tau D} \phi(0) e^{-\tau D} \ .
\end{equation}
Let $\{|E_n\rangle\}$ be a basis of Hamiltonian eigenstates of the Hilbert space. By inserting the identity once, we may write the two-point function in terms of the matrix elements of the field operator as
\begin{align}
     g_\beta(\tau) = \sum_{m,n} e^{-\beta E_n} \langle E_n| e^{\tau H}\phi(0)e^{-\tau H}|E_m\rangle&\langle E_m|\phi(0)|E_n\rangle\\
     & = \sum_{m,n} e^{-\beta E_n +\tau(E_n - E_m)} |\phi_{mn}|^2\ . \nonumber
\end{align}
Here, we have put $\phi_{mn} = \langle E_m |\phi(0)|E_m\rangle$. Now, let $\tau$ and $\kappa$ be real. By a similar manipulation as above
\begin{equation}
    g_\beta(\tau+i\kappa) = \sum_{m,n} e^{-\beta E_n + (\tau+i\kappa)(E_n - E_m)} |\phi_{mn}|^2\ .
\end{equation}
But, clearly we have
\begin{equation}
    \sum_{m,n} e^{-\beta E_n + \tau(E_n - E_m)} |\phi_{mn}|^2 \geq | \sum_{m,n} e^{-\beta E_n + (\tau+i\kappa)(E_n - E_m)} |\phi_{mn}|^2 |\,,
\end{equation}
since all the terms in the first sum are positive and they get multiplied by phases in the second. The last statement is precisely the boundedness condition \eqref{Regge-bound}. 
\smallskip

In addition to the above properties, it is expected that the two-point function $g_\beta(\tau)$ has no singularities in the complex plane, except the one at the origin, $\tau=0$, and its KMS images, $\tau\in\beta\mathbb{Z}$.

\paragraph{Holography} Our discussion so far applies to any conformal field theory. In the following sections, we further assume that the CFT under consideration is holographic and that $\phi$ is a single-trace operator. At leading order in $1/C_T$, the conformal block decomposition \eqref{conformal-block-decomposition} receives contributions from two types of operators only - the double-trace operators $[\phi\phi]_{n,\ell}$ and composites of the stress tensor. We denote the latter schematically by $T_{\mu\nu}^n$ or $[T_{\mu\nu}\dots T_{\rho\sigma}]$. The one-point coefficients $a_\Delta$ corresponding to multi-stress tensors are obtained by analyzing the Klein-Gordon equation on the planar AdS-Schwarzschild black hole in a near boundary expansion, \cite{Fitzpatrick:2019zqz}. Since these coefficients play an important role in our analysis, we give more details about them in Section \ref{SS:Multi-stress tensor OPE coefficients}.

\paragraph{Example: Generalized free field} An example of a thermal two-point function that can be computed exactly is that of the generalized free field (GFF) $\phi$. The GFF two-point function is given by the sum of thermal images of the identity exchange conformal block. At zero spatial separation it can be written in terms of the Hurwitz zeta function $\zeta_H$,
\begin{equation}\label{GFF-two-point-function}
    g^{\text{GFF}}_{\Delta_\phi}(\tau) = \sum_{n=-\infty}^\infty \frac{1}{|\tau+n\beta|^{2\Delta_\phi}} = \frac{1}{\beta^{2\Delta_\phi}} \left( \zeta_H\left(2\Delta_\phi,\frac{\tau}{\beta}\right) + \zeta_H\left(2\Delta_\phi,1-\frac{\tau}{\beta}\right)\right)\ .
\end{equation}
Here, $\Delta_\phi$ denotes the scaling dimension of $\phi$. The operators that appear in the $\phi\times\phi$ OPE in
\eqref{GFF-two-point-function} are the identity operator and the double traces $[\phi\phi]_{n,\ell}$ - see e.g. \cite{Iliesiu:2018fao} for the corresponding coefficients $b_{[\phi\phi]_{n,\ell}}$. The one-point coefficients summed over spins, that appear in \eqref{conformal-block-decomposition-zero-x}, read
\begin{equation}\label{GFF-one-pt-coefficients}
    a^{\text{GFF}}_{[\phi\phi]_{2\Delta_\phi+2m}} = \frac{2 (2m+1)_{2\Delta_\phi - 1}}{\Gamma(2\Delta_\phi)}\zeta(2\Delta_\phi + 2m)\ .
\end{equation}
We have listed those properties of thermal two-point functions that will play a role in the analysis of this work. Additional interesting properties of thermal holographic correlators  include long time behavior, which is controlled by the quasinormal mode frequencies (see e.g. \cite{Horowitz:1999jd,Cardoso:2001bb,Birmingham:2001pj,Policastro:2002se,Nunez:2003eq,Kovtun:2005ev,Brigante:2007nu,Festuccia:2008zx,Buchel:2009tt,Buchel:2009sk,deBoer:2009pn,Berti:2009kk,Kaminski:2009dh,Witczak-Krempa:2012qgh,Emparan:2015rva,Fuini:2016qsc,Grozdanov:2016vgg,Grozdanov:2018gfx,Grozdanov:2019uhi,Jansen:2020hfd,Jeong:2022luo,Dodelson:2024atp,Grozdanov:2024wgo,DeLescluze:2025jqx,Dodelson:2025rng}), pole skipping
(see e.g. \cite{Grozdanov:2017ajz,Blake:2017ris,Haehl:2018izb,Blake:2018leo,Grozdanov:2018kkt,Blake:2019otz,Choi:2020tdj})
and bulk-cone singularities (see e.g. \cite{Hubeny:2006yu,Maldacena:2015iua,Dodelson:2020lal,Dodelson:2023nnr,Chen:2025cee}).

\subsection{Review: multi-stress-tensor OPE coefficients}
\label{SS:Multi-stress tensor OPE coefficients}

\begin{figure}
    \centering
    \includegraphics[width=0.7\linewidth]{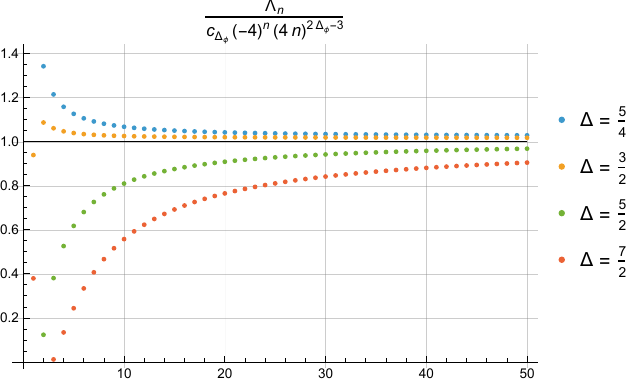}
    \caption{Ratio of the exact stress-tensor coefficients $\Lambda_n$ to their asymptotics for different values of $\Delta_\phi$ and $\alpha_0 =1$ in $d$=4.}
    \label{plot-stress-tensor-coefs}
\end{figure}

Let us separate the holographic two-point function $g_\beta(\tau)$ into three pieces corresponding to the contributions of the identity, double-trace operators and multi-stress tensors, in the conformal block decomposition,
\begin{equation}\label{pieces-two-point-function}
    g_\beta(\tau) = |\tau|^{-2\Delta_\phi} + g_{[\phi\phi]}(\tau) + g_T(\tau)\ . 
\end{equation}
In the leading order in $1/C_T$, the operators do not receive anomalous dimensions. In the following, we will set the number of spacetime dimensions to $d=4$, although conceptually the discussion will remain the same for other values of $d$. The scaling dimensions of double-trace operators take the form $\Delta = 2\Delta_\phi + 2m$, $m\in\mathbb{N}_0$, while the dimensions of multi-stress tensors are $\Delta = 4n$, with $n\in\mathbb{N}$. In this work, unless stated otherwise, we assume that $\Delta_\phi$ is not an integer. Therefore, the double-trace and the stress-tensor contributions in the decomposition \eqref{conformal-block-decomposition-zero-x} do not mix. The conformal block decomposition of the stress-tensor sector will be denoted by 
\begin{equation}\label{multi-stress-tensor-part}
    g_T(\tau) = |\tau|^{-2\Delta_\phi} \sum_{n=1}^\infty \Lambda_n \left(\frac{\tau}{\beta}\right)^{4n}\ .
\end{equation}
The coefficients $\Lambda_n$ may be computed exactly by solving the wave equation on the planar AdS-Schwarzschild black hole, see \cite{Fitzpatrick:2019zqz,Ceplak:2024bja}. The first couple of terms in the expansion read
\begin{align}
    g_T(\tau) = \frac{1}{|\tau|^{2\Delta_\phi}} \Bigg( &\frac{\pi^4\Delta_\phi}{40}\left(\frac{\tau}{\beta}\right)^4\\
    & + \frac{\pi^8\Delta_\phi \left(63\Delta_\phi^4-413\Delta_\phi^3+672\Delta_\phi^2-88\Delta_\phi+144\right)}{201600(\Delta_\phi-4)(\Delta_\phi-3)(\Delta_\phi-2)}\left(\frac{\tau}{\beta}\right)^8 + \dots \Bigg)\ .\nonumber
\end{align}
The most important property of $\Lambda_n$-s for our analysis is their behavior at large $n$. The latter was computed in \cite{Ceplak:2024bja},
\begin{equation}\label{asymptotic-stress-tensor-coefficients}
    \Lambda_n \sim \Lambda_n^0 = c_{\Delta_\phi} \frac{(4n)^{2\Delta_\phi-3}}{\left(\frac{1}{\sqrt{2}}\right)^{4n} e^{i\pi n}}\,, \qquad c_{\Delta_\phi} = \frac{16\alpha_0}{5}\, \frac{\pi\Delta_\phi(\Delta_\phi-1)}{\sin(\pi\Delta_\phi)} \frac{\Delta_\phi}{\Gamma(2\Delta_\phi+3/2)} \ .
\end{equation}
Here, $\alpha_0 = \alpha_0(\Delta_\phi)$ is a numerical coefficient that is  very close to one for $\Delta_\phi$ which is not too large. Figure \ref{plot-stress-tensor-coefs} shows the ratio of the exact and asymptotic stress-tensor coefficients for various values of $\Delta_\phi$. 

\subsection{Sum over images}
\label{SS:Sum over images}

As we have mentioned, the aim of this section is to construct a tentative two-point function $g_\beta(\tau)$ that satisfies all the properties reviewed in Section \ref{SS:Properties of two-point functions} and whose multi-stress-tensor content is given by the asymptotic formula \eqref{asymptotic-stress-tensor-coefficients}. Since \eqref{asymptotic-stress-tensor-coefficients} only agrees with actual holographic multi-stress-tensor coefficients at large $n$, the resulting function $g_\beta(\tau)$ should not be thought of as an exact holographic correlator, but is expected to capture important features of the latter.
\smallskip

We will proceed in two stages. By resumming the asymptotic coefficients \eqref{asymptotic-stress-tensor-coefficients}, we will obtain a tentative stress-tensor sector\footnote{ We will sometimes include the contribution of the identity operator in the stress-tensor sector, and sometimes not -- hopefully this will be clear from the formulas and the context.} of the two-point function, denoted $g_T(\tau)$. The sum over multi-stress-tensor blocks has a finite radius of convergence, but the function $g_T(\tau)$ may be analytically continued beyond the region of convergence of the sum. Next, using the method of images we will generate the function
\begin{equation}\label{g-images}
    g^{\text{images}}_\beta(\tau) = \sum_{m=-\infty}^\infty \left(\frac{1}{|\tau+m\beta|^{2\Delta_\phi}} + g_T(|\tau+m\beta|)\right)\ .
\end{equation}
By construction, this function is KMS-invariant and decomposes into identity, double-trace and stress-tensor sectors. Moreover, the stress-tensor sector of \eqref{g-images} receives contributions only from the zeroth image $m=0$ and is given by $g_T(\tau)$. However, as we will see, the function $g^{\text{images}}_\beta(\tau)$ has singularities within the vertical strip, at 
\begin{equation}\label{singularities}
    \tau_{\pm} = \frac{1\pm i}{2}\beta\ .
\end{equation}
(Consequently, $g^{\text{images}}_\beta(\tau)$ also violates the boundedness condition discussed in Section \ref{SS:Properties of two-point functions}). This brings us to the second stage of the construction, which is concerned with canceling the singularities  of $g^{\text{images}}_\beta(\tau)$ at \eqref{singularities}.

To this end, one can uniquely isolate the {\it singular} part of $g^{\text{images}}_\beta(\tau)$ consisting of the negative-power terms in the Laurent expansion at each of the singularities \eqref{singularities}. This allows us to express $g^{\text{images}}_\beta(\tau)$ as the sum of singular parts centered at \eqref{singularities} and the remaining regular piece
\begin{equation}\label{singular-regular}
    g^{\text{images}}_\beta(\tau) \equiv g^{\text{images}}_{\text{reg}}(\tau) + g^{\text{images}}_{\text{sing}}(\tau) \ .
\end{equation}
Note that $g^{\text{images}}_{\text{reg}}(\tau)$ is regular at $\tau_\pm$, but still contains the KMS poles at $\tau=0,\beta$. Let $K$ be the order of the pole of $g^{\text{images}}_\beta(\tau)$ at $\tau_+$. By KMS-invariance, the pole at $\tau_-$ is of the same order and the Laurent coefficients in expansions around $\tau_+$ and $\tau_-$ coincide up to a sign. Therefore, the singular piece takes the form
\begin{equation}\label{singular-piece}
    g^{\text{images}}_{\text{sing}}(\tau) = \sum_{n=1}^K \left(\frac{r_n}{(\tau - \tau_+)^n} + \frac{(-1)^n r_n}{(\tau - \tau_-)^n}\right)\ .
\end{equation}
We would like to drop the singular piece from $g^{\text{images}}_\beta(\tau)$, but this would violate the KMS condition, since \eqref{g-images} respects KMS symmetry, while \eqref{singular-piece} does not. Instead, we shall subtract from \eqref{singular-regular} the singular part summed over thermal images:
\begin{equation}\label{gbeta-definition}
    g_\beta(\tau) \equiv g^{\text{images}}_\beta(\tau) - \sum_{m=-\infty}^{\infty} g^{\text{images}}_{\text{sing}}(\tau+m\beta) + c \,,
\end{equation}
where $c$ is a constant defined below. This is our definition of the {\it asymptotic model} for the thermal two-point function. Crucially, one shows that the second term in \eqref{gbeta-definition} takes the form of a double-trace sector contribution -- it is an even, KMS-invariant meromorphic function which is regular at points $\tau=0,\beta$. The asymptotic model two-point function $g_\beta(\tau)$ satisfies the KMS condition, has the  stress-tensor content \eqref{asymptotic-stress-tensor-coefficients} and is free of singularities on the vertical strip, except for the OPE pole at $\tau=0$ and its KMS image at $\tau=\beta$. In the following, we shall also verify a weaker form of the boundedness condition \eqref{Regge-bound}, namely the absence of singularities at Im$\,\tau\to\infty$. The constant $c$ in \eqref{gbeta-definition} is chosen such that
\begin{equation}\label{zero-imaginary-infinity}
    \lim_{\text{Im}\, \tau\to\infty} g_\beta(\tau) = 0\ .
\end{equation}
This is a natural property to impose on a two-point function, since the one-point function $\langle\phi\rangle_\beta$ is expected to vanish in holography to leading order in $1/C_T$ expansion.\footnote{ See also \cite{Chai:2020zgq,Komargodski:2024zmt} for recent discussion of symmetry breaking at finite temperature.}

One might also ask whether \eqref{gbeta-definition} is the unique function satisfying the above properties. Although we will not attempt a rigorous proof, there is substantial evidence supporting this. In later sections, we show that the same function arises from the KMS sum rules of \cite{Marchetto:2023xap}, solved by using Borel resummation.
\smallskip

We proceed to illustrate the above procedure explicitly on the example of the scalar field of conformal dimension $\Delta_\phi = 3/2$. The rest of this subsection is dedicated to the construction of $g^{\text{images}}_\beta(\tau)$, while the cancellation of singularities and computation of $g_\beta(\tau)$ is carried out in the next. Other half-integer dimensions are discussed at the end of Section \ref{SS:The asymptotic model}. From now on, we set $\beta=1$.
\smallskip

For $\Delta_\phi=3/2$, the asymptotic multi-stress-tensor coefficients read
\begin{equation}\label{Lambda-n-32}
    \Delta_\phi=\frac32\, : \quad \Lambda_n^0 = - \frac{96\sqrt{\pi}}{175} \alpha_0 (-4)^n\ .
\end{equation}
Summing the stress-tensor sector according to \eqref{multi-stress-tensor-part}, we obtain
\begin{equation}\label{zeroth-approximation-gT}
    g_T(\tau) = - \frac{96\sqrt{\pi} \alpha_0}{175}\, \tau^{-3} \sum_{n=1}^\infty \left(\sqrt{2}e^{-\frac{i\pi}{4}}\tau\right)^{4n} = \frac{384\sqrt{\pi}\alpha_0}{175} \frac{\tau}{1+4\tau^4}\ .
\end{equation}
We have performed the infinite sum within its radius of convergence, $|\tau|\leq1/\sqrt{2}$. We regard the right-hand side of \eqref{zeroth-approximation-gT} as the definition of $g_T(\tau)$ beyond the radius of convergence of the sum. Next, we add the identity exchange contribution and perform the sum over thermal images
\begin{align}
    \Delta_\phi=\frac32\, : \quad g^{\text{images}}_\beta(\tau) &  =\sum_{m=-\infty}^\infty \left(\frac{1}{|\tau+m|^3} + \frac{384\sqrt{\pi} \alpha_0 |\tau+m|}{175\left(1+4(\tau+m)^4\right)}\right)\nonumber\\
   & =  g^{\text{GFF}}_{\Delta_\phi=3/2}(\tau) - \frac{96 i \sqrt{\pi} \alpha_0}{175} \left(\frac{1}{\tau - \frac{1+i}{2}} - \frac{1}{\tau-\frac{1-i}{2}}\right)\ . \label{g0-two-point}
\end{align}
The two terms in the second line are image sums of the two terms in the first, respectively. The first sum is simply the two-point function of the generalized free field with $\Delta_\phi=3/2$. The second sum is exactly evaluated in terms of elementary functions using the identity
\begin{equation}
   \sum_{m=-\infty}^\infty \frac{|\tau+m|}{1 + 4(\tau+m)^4} = \frac{1}{2\left(1-2\tau+2\tau^2\right)}\ .
\end{equation}
By construction, \eqref{g0-two-point} is KMS-invariant. Moreover, it can be seen that only the zeroth image contributes to the multi-stress-tensor content of $g^{\text{images}}_\beta$. Hence, this multi-stress-tensor content is given by $g_T$.

As already mentioned, we consider the function \eqref{g0-two-point} inside the vertical strip. In this domain there are four singularities: the OPE pole at $\tau=0$ and its KMS image at $\tau=1$, together with two further simple poles at $\tau=(1\pm i)/2$. The two terms in \eqref{g0-two-point} correspond to what we call regular and singular pieces in \eqref{singular-regular}.

\subsection{The asymptotic model}
\label{SS:The asymptotic model}

\begin{figure}
    \centering
    \begin{minipage}{0.49\textwidth}
         \centering
        \includegraphics[width=0.8\textwidth]{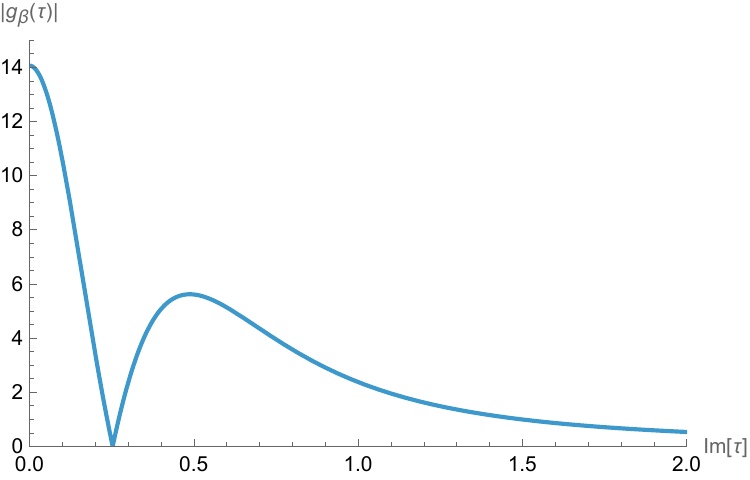}
    \end{minipage}
    \begin{minipage}{0.49\textwidth}
        \centering
        \includegraphics[width=0.8\textwidth]{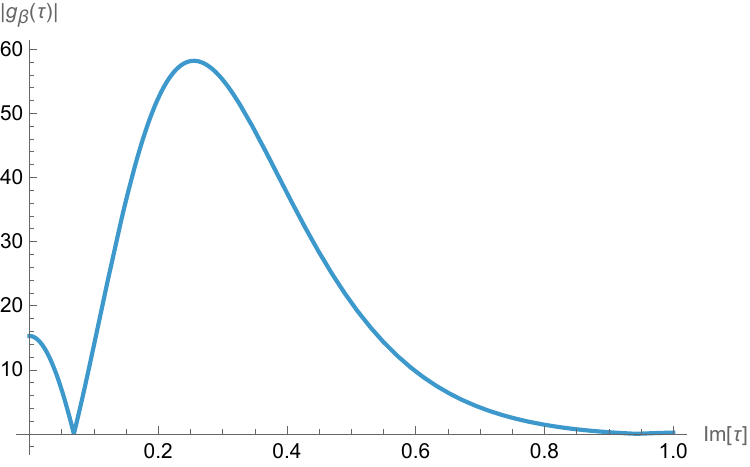}
    \end{minipage}
        \caption{Modulus of $g_\beta(\tau)$ along the line $\Re\tau=\frac12$ for $\Delta_\phi=3/2$ and $\Delta_\phi=5/2$}
    \label{plot-Regge-4d-32}
\end{figure}

Having obtained $g_\beta^{\text{images}}(\tau)$ and identified its regular and singular parts, it remains to perform the subtraction \eqref{gbeta-definition}. To this end, we use the identity
\begin{equation}\label{pole-sum-of-images}
   \frac{i}{2\pi} \sum_{n=-\infty}^\infty \left(\frac{1}{\tau+n - \frac{1+i}{2}} - \frac{1}{\tau+n-\frac{1-i}{2}}\right)= \frac{1}{1+e^{\pi(2i\tau+1)}} - \frac{1}{1 + e^{\pi(2i\tau-1)}}\ .
\end{equation}
We shall denote the right hand side of \eqref{pole-sum-of-images} by $h(\tau)$. This gives us the final result for the asymptotic model
\begin{align}\label{g0-corrected}
    \Delta_\phi=\frac32\, : \quad g_\beta(\tau) =  g^{\text{GFF}}_{\Delta_\phi=3/2}(\tau) & - \frac{96 i \sqrt{\pi} \alpha_0}{175} \left(\frac{1}{\tau - \frac{1+i}{2}} - \frac{1}{\tau-\frac{1-i}{2}}\right) \\
    &+ \frac{192\pi^{3/2}\alpha_0}{175} \left(\frac{1}{1+e^{\pi(2i\tau+1)}} - \frac{1}{1 + e^{\pi(2i\tau-1)}} \right) \ .\nonumber
\end{align}
The expression \eqref{g0-corrected} applies in the strip $0\leq \Re\tau\leq 1$. As advocated in our general discussion, $g_\beta(\tau)$ satisfies all of the properties listed in Section \ref{SS:Properties of two-point functions}. In particular, it is free of singularities apart from ones at $\tau=0,1$ and bounded as $\text{Im}\,\tau\to\infty$. Note that we may try to add to $g_\beta(\tau)$ a finite linear combination of functions
\begin{equation}\label{functions-fs}
    f_s (\tau) = \cos(2\pi s\tau)\,, \qquad s\in\mathbb{N}_0\,,
\end{equation}
which are KMS-invariant and decompose into double-trace conformal blocks.\footnote{For $\Delta_\phi=3/2$, multi-stress-tensor blocks are given by odd powers of $\tau$ and double-trace blocks by even powers.} While this would not change the analytic structure of \eqref{g0-corrected}, any such linear combination diverges as $\text{Im}\, \tau\to\infty$ and therefore violates the boundedness condition \eqref{Regge-bound}. This is with the exception of the $s=0$ case, that is, a constant function. The value of this additive constant is fixed by the requirement \eqref{zero-imaginary-infinity}, which is indeed satisfied by \eqref{g0-corrected}. One property that is not automatically built in our construction of \eqref{g0-corrected} is the boundedness condition \eqref{Regge-bound}, which is stronger than the absence of singularities.
However, it can be verified that \eqref{Regge-bound} {\it is} indeed satisfied. For instance, the plot on the left in Figure \ref{plot-Regge-4d-32} shows the behavior of $|g_\beta(\tau)|$ along the line $\Re\tau=1/2$. One can draw similar plots for other values of $\kappa$ in \eqref{Regge-bound} and observe that the boundedness condition \eqref{Regge-bound} always holds in the case $\Delta_\phi=3/2$.
It is not necessarily the case for other values of $\Delta_\phi$.

The one-point coefficients of the asymptotic model can readily be computed. They are given by
\begin{align}\label{AM-double-trace-coefficients-32}
    \Delta_\phi=\frac32\, : \quad a^{\text{AM}}_{[\phi\phi]_{2\Delta_\phi+2m}} & = a^{\text{GFF}}_{[\phi\phi]_{2\Delta_\phi+2m}}\\
    & +\frac{192\sqrt{\pi}\alpha_0}{175}\left((-1)^{\lfloor m/2 \rfloor} 2^m + (-1)^m (2 \pi )^{2 m+1} \frac{\text{Li}_{-2 m}\left(-e^{\pi }\right)}{(2m)!}\right)\,,\nonumber
\end{align}
for $m\in\mathbb{N}$, and 
\begin{equation}
    \Delta_\phi=\frac32\, : \quad a^{\text{AM}}_{[\phi\phi]_{2\Delta_\phi}} = 2\zeta(3) + \frac{192\sqrt{\pi}\,\alpha_0}{175} \left(1 - 2\pi \tanh\left(\frac{\pi}{2}\right)\right)\,,
\end{equation}
for \( m = 0 \). Here, \( \lfloor x \rfloor \) denotes the integer part of $x$ and Li$_s(x)$ the polylogarithm.

\paragraph{Other values of $\Delta_\phi$} We end this subsection by discussing the above construction for other values of the conformal dimension $\Delta_\phi$. Whenever the latter takes the form $\Delta_\phi = p+1/2$, with $p\in\mathbb{N}$, the asymptotic model can be constructed explicitly. Firstly, resumming the asymptotic stress-tensor data \eqref{asymptotic-stress-tensor-coefficients}, we obtain the function $g_T(\tau)$, which is a rational function. The sum over thermal images $g_\beta^{\text{images}}(\tau)$ is then given by a combination of rational and polygamma functions. Considering it on the vertical strip, this function has KMS poles at $\tau=0,1$, as well as two poles at $(1\pm i)/2$. The order of the latter depends on the value of $\Delta_\phi$ and equals $2p-1$. Finally, in order to pass to the asymptotic model we perform the sum \eqref{singular-piece} - this is given by a linear combination of powers $\{h^0,h^1,h^2,\dots,h^{2p-1}\}$ of the function $h(\tau)$ given in \eqref{pole-sum-of-images}. \footnote{An alternative point of view of what we have described is the following. After obtaining the function $g^{\text{images}}_\beta(\tau)$, we may try to cancel its poles by a linear combination of $f_s(\tau)$. Since $f_s(\tau)$ have no singularities, we need an infinite linear combination. In particular, the function \eqref{pole-sum-of-images} arises as the formal sum
\begin{equation}
   h(\tau) = \sum_{s=-\infty}^\infty (-1)^s e^{\pi |s|} \cos(2\pi s \tau) \ . \nonumber
\end{equation}
For half-integer $\Delta_\phi$, there is a unique finite linear combination of powers of $h$ that can be added to $g^{\text{images}}_\beta(\tau)$ in order to cancel singularities within the vertical strip.}



Note that the boundedness condition \eqref{Regge-bound} can be violated in the asymptotic model depending on the value of $\Delta_\phi$. For example, this is the case for $\Delta_\phi = 5/2$, as shown in right plot of Figure \ref{plot-Regge-4d-32}. Indeed, in contrast to the plot on the left, the maximum of the function is above the value at Im$\,\tau=0$. 
It is expected that the condition \eqref{Regge-bound} is sensitive to fine details of the stress-tensor OPE coefficients (because it is a property of the exact thermal state), and therefore it is not captured by the asymptotic model. 
In this respect, the violation of the boundedness condition in Figure \ref{plot-Regge-4d-32} for $\Delta_\phi=5/2$ highlights the inaccuracy of the model.

\subsection{Refinements of the model}
\label{SS:Refinements of the model}

While it is not a subject that we will pursue further in the present work, in this subsection we briefly discuss the important question of how one might try to systematically refine our asymptotic model and make it closer to the actual holographic two-point function.
\smallskip

Until now our analysis only made use of the asymptotic form of multi-stress-tensor one-point coefficients \eqref{asymptotic-stress-tensor-coefficients}. It is natural to expect that by including more information about the exact values of these coefficients into the model, one gets closer to the holographic two-point function. Here we discuss two possible approaches to inputting further multi-stress-tensor data. The first one, similar to what was considered in \cite{Parisini:2023nbd}, consists of replacing, one by one, the low-lying coefficients $\Lambda_n^0$ by their exact values $\Lambda_n$. After replacing a finite number $N$ of coefficients, one may try to construct the corresponding asymptotic model along the same lines as above. Let us denote the stress-tensor sector that we start with by
\begin{equation}\label{Skenderis-like-correction-gT}
    g_T^{(N)}(\tau) = g_T(\tau) + |\tau|^{-2\Delta_\phi} \sum_{n=1}^N \delta\Lam_n\, \tau^{4n}\,, \qquad \delta\Lam_n=\Lambda_n - \Lambda_n^0\ .
\end{equation}
The first step of the construction consists in taking the sum over thermal images. However, we face an immediate difficulty in the fact that the sum over images of the second term in \eqref{Skenderis-like-correction-gT} diverges whenever $4N>2\Delta_\phi$. Following \cite{Parisini:2023nbd}, one can make the sum finite using $\zeta$-function regularization. To give more details, consider the generically divergent (for $2n>\Delta_\phi$) sum over images of a single conformal block
\begin{equation}\label{sum-of-images-block}
    B_n = \delta\Lam_n \sum_{m=-\infty}^\infty |\tau+m|^{4n-2\Delta_\phi}\ .
\end{equation}
To make sense of the sum, one uses the analytic continuation of the formula
\begin{equation}\label{zeta-function-regularisation}
\sum_{m=1}^\infty |\tau+m|^\gamma = \zeta_H(-\gamma,1+\tau)\,,
\end{equation}
that is valid for $\gamma<-1$. The right-hand side of \eqref{zeta-function-regularisation} is well defined for all $\gamma\neq-1$. Plugging this into \eqref{sum-of-images-block}, we get that each OPE coefficient correction $\delta\Lambda_n$ contributes
\begin{equation}
    B_n^{\text{reg}} = \delta\Lam_n \Big( \zeta_H(2\Delta_\phi-4n,\tau) + \zeta_H(2\Delta_\phi-4n,1-\tau)\Big) \equiv \delta\Lam_n\,g^{\text{GFF}}_{\Delta_\phi-2n}(\tau)\,,
\end{equation}
to the sum over images. Therefore, we obtain
\begin{equation}
    g_\beta^{\text{images},(N)}(\tau) = g_\beta^{\text{images}}(\tau) + \sum_{n=1}^N \delta\Lam_n\, g^{\text{GFF}}_{\Delta_\phi-2n}(\tau)\ .
\end{equation}
The function added to $g_\beta^{\text{images}}(\tau)$ is entire for half-integer values of $\Delta_\phi$ that we consider in this work. Therefore, the singular part of $g_\beta^{\text{images},(N)}(\tau)$ is the same as that of $g_\beta^{\text{images}}(\tau)$. This leads to the refinement of the asymptotic model
\begin{equation}\label{low-operator-corrected-CAM}
    g_\beta^{(N)}(\tau) = g_\beta(\tau) + \sum_{n=1}^N \delta\Lam_n\, g^{\text{GFF}}_{\Delta_\phi-2n}(\tau)\ .
\end{equation}
Let us argue that \eqref{low-operator-corrected-CAM} is actually not a good way of adjusting the asymptotic model. Assume that $\Delta_\phi = p+1/2$, where $p$ is a positive integer. Then, for $2n>\Delta_\phi$, the function $g^{\text{GFF}}_{\Delta_\phi-2n}(\tau)$ is given in terms of the Bernoulli polynomials of degree $4n-2p$
\begin{align}
    g^{\text{GFF}}_{\Delta_\phi-2n}(\tau) &=  \zeta_H(2\Delta_\phi-4n,\tau) + \zeta_H(2\Delta_\phi-4n,1-\tau) \\
    \nonumber &= -\frac{1}{4n+2\Delta_\phi+1}\left(B_{4n+2\Delta_\phi+1}(\tau) + B_{4n+2\Delta_\phi+1}(1-\tau)\right)\,,
\end{align}
and hence it diverges as Im$\,\tau\to\infty$. Therefore, for a large enough $N$, the corrections \eqref{low-operator-corrected-CAM} violate the analytic structure that we require of the two-point function. Furthermore, as will be discussed in Section \ref{SS:Application to the refined model}, the double-trace coefficients associated with the refined asymptotic model do not converge to a limiting value as $N\to\infty$, see in particular the equations \eqref{Borel_corrections} and \eqref{dominant-correction-am}.

Having ruled out \eqref{low-operator-corrected-CAM} as a systematic procedure that can be extended infinitely in $N$, let us note that a certain finite number of terms in \eqref{low-operator-corrected-CAM} may be added while preserving the analytic structure of the two-point function. To see this, note that for $2n<\Delta_\phi-2$, the function $g^{\text{GFF}}_{\Delta_\phi-2n}(\tau)$ decays as Im$\,\tau\to\infty$ and is free from poles on the vertical strip. It follows that adding the finite number of corrections \eqref{low-operator-corrected-CAM} with $n=1,2,\dots,\lfloor p/2\rfloor -1$ does not alter the analytic structure of $g_\beta(\tau)$. The resulting correlator with $N=\lfloor p/2\rfloor-1$ has the exact one-point coefficients $\Lambda_1,\dots,\Lambda_{\lfloor p/2\rfloor -1}$ for the first $\lfloor p/2\rfloor -1$ composites of the stress-tensor and asymptotic coefficients \eqref{asymptotic-stress-tensor-coefficients} for the higher composites.
\smallskip

To make progress beyond the above and obtain a refinement scheme that would converge to the actual holographic two-point function, one may try to modify all multi-stress-tensor coefficients at once in a large $n$-expansion such as
\begin{equation}\label{large-n-corrections}
    \Lambda_n^{[M]} = \Lambda_n^0 \sum_{s=0}^M \frac{\alpha_s}{n^s}\,, \quad n\to\infty\ .
\end{equation}
It remains to be seen what precise form of the large $n$ expansion should be used (e.g. one can try expanding in powers of $n^{-a}$ for any $a>0$) and if the resulting correction procedure converges. This task is challenging primarily because the exact multi-stress coefficients $\Lambda_n$ are computed recursively with increasing $n$, making the large-$n$ data not easily accessible. We defer the discussion of large-$n$ refinements of the asymptotic model to the concluding section.

\section{Sum rules}
\label{S:Sum rules}

In this section, we analyze the sum rules arising from the KMS condition and show how they can be solved. After reviewing the sum rules, which take the form of an infinite system of linear equations, in the first subsection, we go on to derive the homogeneous and a particular solution in the next two subsections. The particular solution requires regularization, which is performed using the Borel summation method. We proceed to apply the procedure to the GFF sum rules, recovering the correct OPE coefficients \eqref{GFF-one-pt-coefficients}. This is followed by solving the sum rules that arise from the asymptotic multi-stress-tensor CFT data \eqref{asymptotic-stress-tensor-coefficients} ({\it asymptotic holographic sum rules} for short) and showing that the solution coincides with the results given by the asymptotic model of Section \ref{S:Toy models of holography}. The section ends with a discussion of how these solutions are affected by refinements of the input CFT data (the multi-stress-tensor one-point  coefficients).

\subsection{Review: KMS sum rules}

The KMS condition ensures that the thermal correlation functions exhibit periodicity along the thermal circle. Any two-point function at finite temperature in Euclidean time satisfies \eqref{KMS-condition}. Let us review the derivation of the explicit sum rules out of this condition for the simplest case of zero spatial separation between the operators\footnote{See \cite{Marchetto:2023xap} for a similar derivation in the more general case of non-zero spatial separation.}. We start from the conformal block decomposition of the two-point function \eqref{conformal-block-decomposition-zero-x}. By substituting this expansion into the KMS equation \eqref{KMS-condition} and applying the binomial theorem in the form
\begin{equation}
    \left(\frac{\beta}{2} \pm \delta\tau\right)^{\Delta-2\Delta_\phi} = \sum_{\tilde k=0}   \frac{\Gamma(\Delta-2\Delta_\phi+1)}{\Gamma(\Delta-2\Delta_\phi-\tilde k+1)\Gamma(\tilde k+1)}\left(\frac{\beta}{2}\right)^{\Delta-2\Delta_\phi-\tilde k}(\pm\delta\tau)^{\tilde k}\,,
\end{equation}
we obtain:
\begin{equation}
\label{kms1}
    \sum_{\Delta} \frac{a_\Delta}{2^\Delta} \frac{\Gamma(\Delta - 2\Delta_\phi+1)}{\Gamma(\Delta - 2\Delta_\phi - 2k )} = 0\,, \qquad k\in\mathbb{N}_0\,, 
\end{equation}
where we used \eqref{KMS-condition}, which ensures that only odd values of $\tilde k =2 k +1$ yield nontrivial condition \eqref{kms1}. The contribution from the identity operator $\mathbbm {1}$ can be separated from the rest of the sum, leading to the sum rules
\begin{equation}\label{sum_rules}
    \sum_{\Delta} \frac{a_\Delta}{2^\Delta} \frac{\Gamma(\Delta - 2\Delta_\phi+1)}{\Gamma(\Delta - 2\Delta_\phi - 2k )}  = \frac{\Gamma(2\Delta_\phi+2k+1)}{\Gamma(2\Delta_\phi)}\,, \quad k\in\mathbb{N}_0\ . 
\end{equation}
An alternative perspective on equations \eqref{sum_rules} is that expanding the two-point function around the KMS-symmetric point, $\tau=\frac{\beta}{2}$, is equivalent to imposing the conditions:
\begin{equation}
    \partial_t^{2k+1} \left.g(\tau,0)\right|_{\tau=\frac{\beta}{2}} = 0\,, \qquad k \in \mathbb{N}_0\ .
\end{equation}
The sum rules \eqref{sum_rules} constitute an infinite set of linear equations. Let us note that numerical approaches to solving \eqref{sum_rules} based on truncating the system meet with difficulties - such approaches are best suited for systems where coefficients $a_\Delta$ are assumed to be positive. However, one cannot justify this assumption in generic CFTs. In this work, we focus instead on analytic approaches, showing that despite the complexity of the sum rules, some solutions can be found exactly. Specifically, we will be interested in the sum rules in the context of holographic CFTs, where the $\phi\times\phi$ OPE, in the leading order, consists solely of the identity, double-trace and multi-stress-tensor operators
\begin{equation}
    \phi\times\phi \sim \mathbbm{1} + [\phi\phi]_{2\Delta_\phi+2m} + [T_{\mu\nu}]^{n}\ .
\end{equation}
In this case, the sum rules take the form
\begin{equation}\label{hol_sum_rules}
    \sum_{m > k} \frac{{a_{[\phi\phi]}}_{2\Delta_\phi+2m}}{2^{2\Delta_\phi+2m}}   \frac{\Gamma(2m+1)}{\Gamma(2(m-k))}  = \frac{\Gamma(2\Delta_\phi+2k+1)}{\Gamma(2\Delta_\phi)} - \sum_{n=1
}^{\infty} \frac{\Lambda_n}{2^{dn}}   \frac{\Gamma(dn - 2\Delta_\phi+1)}{\Gamma(dn-2\Delta_\phi-2k)}\ .
\end{equation}
Although the stress-tensor part in \eqref{hol_sum_rules} involves gamma functions that can develop poles for certain values of $n$, all such singularities cancel between the numerator and the denominator.
As a result, each term in the series remains finite and the overall sum rules are well-defined. The coefficients of the double-trace operators are regarded as unknown variables, whereas those corresponding to multi-stress-tensor operators are taken to be known. 
First, we turn to the solution to the homogeneous system of equations, where the right-hand side is equal to zero.

\subsection{Homogeneous equations}\label{Homogeneous equations}

In this subsection, we derive the solutions to the homogeneous equations
\begin{equation}\label{homogeneous-eqns}
    \sum_{m > k} \frac{{a_{[\phi\phi]}}_{2\Delta_\phi+2m}}{2^{2\Delta_\phi+2m}} \frac{\Gamma(2m+1)}{\Gamma(2(m-k))} = 0\,, \qquad k\in\mathbb{N}_0\ .
\end{equation}
As stated in \cite{fedorov2015specificities} for general systems of this kind, the general solution to \eqref{homogeneous-eqns} can be expressed in the form
\begin{equation}
    a_{[\phi\phi]_{2\Delta_\phi+2m}} = \frac{a(-1)^m}{S^{m}(2m)!}\,, \quad m\in \mathbb{N}\,,
\end{equation}
where $a$ is an arbitrary real number and $S$ is a constant {\it characteristic number}. In the case at hand, $S$ is determined from the non-linear equation
\begin{equation}\label{nl_eq}
    \sum_{m=1}^\infty\frac{(-1)^m}{(4S)^m (2m-1)!} = 0\ .
\end{equation}
In order to solve \eqref{nl_eq}, we can construct a characteristic function
\begin{equation}\label{f(x)}
    f(x) = \sum_{m=1}^\infty\frac{(-1)^mx^m}{(2m-1)!} = -\sqrt{x}\sin{\left(\sqrt{x}\right)}\,,
\end{equation}
and find its zeros, $f(x_n)=0$,
\begin{equation}\label{hom_sol}
     x_n = (\pi n)^2, \qquad n \in \mathbb{N}_0\ .
\end{equation}
The corresponding $S_n = \frac{1}{(2\pi n )^2}$ lead to the solutions of the homogeneous system of equations \eqref{homogeneous-eqns}
\begin{equation}\label{homogeneous-solutions}
    a_{[\phi\phi]_{2\Delta_\phi+2m}} = \frac{a(-1)^m (2\pi n)^{2m}}{(2m)!}\,, \qquad m\in \mathbb{N}\,,
\end{equation}
for any positive integer $n$ and a real number $a$. Note that the homogeneous solution \eqref{homogeneous-solutions} corresponds to the freedom to add the function $a\cos(2\pi n \tau)$ to the two-point function $g_\beta(\tau)$.
This is in accord with our discussion around \eqref{functions-fs} in the previous section.

\subsection{Particular solution}\label{Particular solution}

Having solved the homogeneous problem, we will now develop a method to solve the sum rule equations for a general right hand side, denoted by  \( F_k \).
The corresponding equation is given by
\begin{equation}\label{general sum rules}
    \sum_{m > k} \frac{{a_{[\phi\phi]}}_{2\Delta_\phi+2m}}{2^{2\Delta_\phi+2m}} \frac{\Gamma(2m+1)}{\Gamma(2(m-k))} = F_k\,, \quad k\in\mathbb{N}_0\ .
\end{equation}
The key to solving this linear system of equations is to recognize that the matrix is an upper-triangular Toeplitz matrix, allowing for an efficient solution via backward substitution. We will first outline the method by simplifying the equations and deriving a recursive procedure to obtain the solution. Then we will solve the resulting recursion relations. We begin by expressing the system in the standard form:
\begin{equation}\label{linear_system}
   \sum_{m=1}^\infty M_m a_{m+k} = F_k\,, \qquad k \in \mathbb{N}_0\,,
\end{equation}
where we denoted
\begin{equation}\label{notations}
    a_m = \frac{(2m)!\, {a_{[\phi\phi]}}_{2\Delta_\phi+2m}}{2^{2\Delta_\phi+2m}} \quad\text{and}\quad M_m = \frac{1}{(2m-1)!}\ .
\end{equation}
To determine \( a_1 \), add the second equation of \eqref{linear_system}, multiplied by \( p_2 = -M_2 \), to the first equation. Subsequently, the third equation, multiplied by \( p_3 = -M_3 - p_2 M_2 \), is added to the first equation. Repeating this procedure iteratively yields the expression  
\begin{equation}
    a_1 = \sum_{k=0}^\infty p_{k+1}F_k\,,
\end{equation}  
where we define $p_1=1$ and for $k>2$ the $p_k$ satisfy recurrence relations
\begin{equation}\label{recurrence}
    p_k = -M_k - \sum_{n=2}^{k-1} M_{k-n+1}p_n\ .
\end{equation}
We can repeat this procedure for any row, which leads to  
\begin{equation}\label{particular_new}
    a_m = \sum_{k=0}^\infty p_{k+1} F_{m+k-1}\,, \qquad m\in\mathbb{N}\ .
\end{equation}
Now we proceed to solve the recurrence relation \eqref{recurrence}. Constructing two generating functions
\begin{equation}
    P(x) = \sum_{k=2}^\infty p_k x^k, \qquad M(x) = \sum_{k=2}^\infty M_k x^k = \sqrt{x} \sinh \left(\sqrt{x}\right)-x\,,
\end{equation}
and switching the summation limits in \eqref{recurrence} with the change of variables, we can write the identity 
\begin{equation}
    p_2 x^2 + P(x) = -M(x) - M_2 x^2 -\frac{1}{x}P(x) M(x)\ .
\end{equation}
Rearranging, this gives
\begin{equation}
    P(x) = -\frac{M(x)}{1+\frac{M(x)}{x}} = x \left(\sqrt{x}\, \text{csch}\left(\sqrt{x}\right)-1\right)\ .
\end{equation}
By expanding $P(x)$ into a series, we find 
\begin{equation}\label{pk-coefficients}
    p_k = \frac{2 \left( 1 - 2^{2k - 3} \right) B_{2k - 2}}{(2k - 2)!}\,, \qquad k\in\mathbb{N}\,,
\end{equation}
where $B_k$ denote the Bernoulli numbers. In the following, we will find it convenient to use an alternative expression for the coefficients $p_k$ by writing the Bernoulli numbers in terms of the Riemann zeta function,
\begin{equation}\label{pk-zeta-expression}
    p_k = 2(-1)^{k-1} \frac{\zeta(2k-2)}{\pi^{2k-2}}\left(1 - \frac{2}{4^{k-1}}\right)\,, \qquad k\in\mathbb{N}\ .
\end{equation}
In particular, for $k=1$, one uses the analytically continued value $\zeta(0) = -\frac12$. Putting everything together, we arrive at the final result 
\begin{equation}\label{particular_solution}
    a_{[\phi\phi]_{2\Delta_\phi+2m}} = \frac{2^{2\Delta_\phi+2m}}{(2m)!}\sum_{k=0}^{\infty} (-1)^k \frac{\zeta(2k)}{\pi^{2k}} \left(2-4^{1-k}\right) F_{k+m-1}\,, \quad m\in \mathbb{N}\ .
\end{equation}
Note that for the sum to converge, \( F_k \) must satisfy the following asymptotic condition:
\begin{equation}
    F_k = o(\pi^{2k}) \quad \text{as} \quad k \to \infty\ .
\end{equation}
This is particularly significant, as the structure of the KMS sum rules in fact implies the factorial growth $F_k \sim (2k)!$, rendering \eqref{particular_solution} ill-defined. To make sense of the divergent sum, one can employ the Borel summation\footnote{More precisely, a series whose partial sums are bounded by \((nk)! \, C^{k+1}\) for some constant \(C > 0\) and integer $n$ is said to be summable in the sense of Mittag-Leffler.}, which provides a prescription for handling such asymptotic series. To this end, consider the generating series associated with the solution \eqref{particular_new}
\begin{equation}
    a_m(z) \equiv \sum_{k=0}^\infty p_{k+1}\, F_{m+k-1}\, z^k\ .
\end{equation}
In order to regularize the series, we perform the generalized Borel transform
\begin{equation}\label{generalised-Borel-transform}
    \mathcal{B}_m(t^2z) = \sum_{k=0}^{\infty} \frac{p_{k+1}\, F_{m+k-1}\, z^k}{(2k)!}t^{2k}\ .
\end{equation}
Then the Borel summation of $a_m$ is given by
\begin{equation}\label{Borel-resummed-am}
    a_m = a_m(1) \equiv \int^\infty_0 e^{-t}\mathcal{B}_m(t^2)\, dt\ .
\end{equation}
We have slightly abused the notation in denoting both the divergent series \eqref{particular_new} and its resummation \eqref{Borel-resummed-am} by the same letter. Using the rule \eqref{Borel-resummed-am} and the explicit form of the coefficients $p_k$, the particular solution for the general sum rules \eqref{general sum rules} can be expressed as
\begin{align}\label{regularized_solution}
    &a_{[\phi\phi]_{2\Delta_\phi+2m}} =  \frac{2^{2\Delta_\phi+2m}}{(2m)!}\int_0^\infty e^{-t} \sum_{k=0}^{\infty}\frac{(-1)^{k}}{(2k)!} \frac{\zeta(2k)}{\pi^{2k}}\left(2-4^{1-k}\right)F_{k+m-1}\, t^{2k}dt\,, \quad m\in \mathbb{N}\ .
\end{align}
This is the main general formula of the present section. In the next two subsections, we show how \eqref{regularized_solution} can be explicitly evaluated in GFF and asymptotic holography. 

\subsection{Reproducing GFF}\label{Reproducing GFF}

To illustrate one of the possible applications of the  solution \eqref{regularized_solution}, we consider the case of the GFF, where the right hand side is given by 
\begin{equation}
   F_k^{\text{GFF}} = \frac{\Gamma(2\Delta_\phi+2k+1)}{\Gamma(2\Delta_\phi)}\ .
\end{equation}
We work with the notation introduced in \eqref{notations}. According to \eqref{pk-zeta-expression} and \eqref{generalised-Borel-transform}, 
the generalized Borel transform $\mathcal{B}^{\text{GFF}}_m(t^2)$ is
\begin{align}
        \mathcal{B}^{\text{GFF}}_m(t^2) &= \sum_{k=0}^{\infty} (-1)^k \frac{\zeta(2k)}{\pi^{2k}} \left(2-4^{1-k}\right)\frac{\Gamma(2k+2m+2\Delta_\phi-1)}{\Gamma(2\Delta_\phi)\Gamma(2k+1)}t^{2k}\ .
\end{align}
To effectively evaluate this sum, we make use of the following identity: 
\begin{equation}\label{auxiliary_sum}
    \sum_{k=0}^{\infty} (-1)^k\frac{\zeta(2k)}{\pi^{2k}}\left(2-4^{1-k}\right) t^{2k} = t \csch(t)\ .
\end{equation}
This representation allows for a simpler expression for the Borel transform, which can be re-written as
\begin{equation}
    \mathcal{B}^{\text{GFF}}_m(t^2) = \frac{1}{\Gamma(2\Delta_\phi)}\partial_t^{2m+2\Delta_\phi-2}\left(t^{2m+2\Delta_\phi-1}\csch(t)\right)\ .
\end{equation}
Therefore, in the case at hand, the coefficients $a_m$ are evaluated to
\begin{align}\label{GFF_coefs}
     a^{\text{GFF}}_{\Delta_\phi,\,m} &= \frac{1}{\Gamma(2\Delta_\phi)}\int^\infty_0 e^{-t}\, \partial_t^{2m+2\Delta_\phi-2}\left(t^{2m+2\Delta_\phi-1} \csch(t)\right)dt \nonumber\\ 
    &= \frac{1}{\Gamma(2\Delta_\phi)}\int^\infty_0e^{-t}t^{2m+2\Delta_\phi-1} \csch(t)dt \\ \nonumber
    &= 2^{-2m-2\Delta_\phi+1} \zeta (2m + 2\Delta_\phi) \frac{\Gamma (2m + 2\Delta_\phi)}{\Gamma(2\Delta_\phi)}\,, \nonumber
\end{align}
where, to get to the second line, we used integration by parts and the facts that $t\csch(t)-1 \sim t^2$ as $t\to0$ and $\csch(t)$ is bounded for $t\to\infty$. Plugging this into the formula for the double-trace coefficients \eqref{notations}, we find
\begin{equation}
    a^{\text{GFF}}_{[\phi\phi]_{2\Delta_\phi+2m}} = \frac{a^{\text{GFF}}_{\Delta_\phi,\,m} \,2^{2m+2\Delta_\phi}}{\Gamma(2m + 1)} = \frac{2 (2m+1)_{2\Delta_\phi - 1}}{\Gamma(2\Delta_\phi)}
     \zeta(2\Delta_\phi+2m)\ .
\end{equation}
The result coincides with the thermal one-point coefficients of the GFF theory, \eqref{GFF-one-pt-coefficients}.

\subsection{Application to the asymptotic model}
\label{SS:Borel-toy-model}

Now we apply the solution \eqref{regularized_solution} to the holographic sum rules \eqref{hol_sum_rules}. Due to linearity, it suffices to solve equations \eqref{hol_sum_rules} with the right-hand side consisting only of the second, multi-stress-tensor term, as the contribution from the identity is given by the GFF solution. Using the notation introduced in \eqref{notations}, the multi-stress-tensor contribution is expressed as 
\begin{equation}\label{stress rhs}
    F_k = -\sum_{n=1}^\infty\frac{\Lambda_{n}}{2^{4n}}\frac{\Gamma({4n} - 2\Delta_\phi+1)}{\Gamma({4n} - 2k - 2\Delta_\phi)} \ .
\end{equation}
We consider the case where $\Delta_\phi$ is a half-integer. In particular, we analyze the two distinct scenarios separately:  
\begin{equation}\label{conformal_dimensions}
\Delta_\phi = 2r-\frac12 \quad \text{and} \quad \Delta_\phi = 2r+\frac12\,, \quad r \in \mathbb{N}\ .
\end{equation}
As an example, consider first the simplest case, $\Delta_\phi = \frac32$. Explicit expressions for $F_k$ are not available for exact holography, where one knows only a finite number of coefficients $\Lambda_n$. However, if the coefficients are replaced by their large $n$ asymptotics, $F_k$ can be computed. We will denote these numbers by $\hat F_k$. To facilitate the analysis, it is convenient to analyze the coefficients $\hat F_k$ in two separate cases
\begin{align}
        \label{F2k}\Delta_\phi=\frac32\, : \quad \hat F_{2k} & 
    = -\frac{1}{2^{4(k+1)}}\sum_{n=0}^\infty \frac{\Lambda_{n+k+1}^0}{2^{4n}}\frac{\Gamma(4n+4k + 2)}{\Gamma(4n + 1)} \,,\\
    \label{F2k1}\hat F_{2k-1} &
    = -\frac{1}{2^{4(k+1)}}\sum_{n=0}^\infty \frac{\Lambda_{n+k+1}^0}{2^{4n}}\frac{\Gamma(4n+4k + 2)}{\Gamma(4n + 3)}\ .
\end{align}
Explicit expressions for $\hat F_{2k}$ and $\hat F_{2k-1}$ are given in Appendix \ref{identities}, see equation \eqref{Fk-appendix}. As we have seen, the solution for the double-trace coefficients is given by an integral over the Borel-transformed sum
\begin{equation}\label{Borel-transform-holography}
        \mathcal{\hat B}_m(t^2) = \sum_{k=0}^{\infty} (-1)^k \frac{\zeta(2k)}{\pi^{2k}} \left(2-4^{1-k}\right)\frac{\hat F_{k+m-1}}{\Gamma(2k+1)}t^{2k} , \qquad m = 0,1,2,\ldots\ .
\end{equation}
We consider two cases, $\mathcal{\hat B}_{2m}(t^2)$ and $\mathcal{\hat B}_{2m-1}(t^2)$, in turn.

\subsection*{Even $\bf{\mathcal{\hat B}_{2m}}$}

The sum \eqref{Borel-transform-holography} becomes
\begin{align}
     \nonumber \mathcal{\hat B}_{2m}(t^2) &=  \sum_{k=0}^{\infty}  \frac{\zeta(4k)}{\pi^{4k}} \left(2-4^{1-2k}\right)\frac{\hat F_{2k+2m-1}}{\Gamma(4k+1)}t^{4k} \\  &- \sum_{k=1}^{\infty}  \frac{\zeta(4k-2)}{\pi^{4k-2}} \left(2-4^{2-2k}\right)\frac{\hat F_{2k+2m-2}}{\Gamma(4k-1)}t^{4k-2}\ . \label{hat-B2m}
\end{align}
We will analyze each of the two sums separately. Let us denote the first sum by $B_1$. It takes the form
\begin{equation}
\label{b1}
    B_1 = -\sum_{k=0}^{\infty}\sum_{n=0}^\infty  \frac{\zeta(4k)}{\pi^{4k}} \left(2-4^{1-2k}\right)\frac{\Gamma(4n+4k + 4m+2)}{\Gamma(4k+1)\Gamma(4n + 3)}\frac{\Lambda_{n+k+m+1}^0}{2^{4n+4k+4m+4}} t^{4k}\ .
\end{equation}
We now use the explicit form of the asymptotic $\Lambda^0_n = c_{3/2}(-4)^n$ from \eqref{Lambda-n-32}, as this greatly simplifies the analysis. 
Then, \eqref{b1} becomes
\begin{align}
    B_1 =  \frac{c_{3/2}}{2}(-4)^{-m}\sum_{n=0}^\infty  \frac{(-4)^{-n}}{\Gamma(4n + 3)}\sum_{k=0}^{\infty}\frac{\zeta(4k)}{(-4\pi^4)^{k}} \left(1-\frac{2}{4^{2k}}\right)\frac{\Gamma(4n+4k + 4m+2)}{\Gamma(4k+1)} t^{4k}\ .
\end{align}
It is convenient to define
\begin{equation}\label{S1(t)-definition}
    S_1(t) \equiv \sum_{k=0}^{\infty}\frac{\zeta(4k)}{(-4\pi^4)^{k}} \left(1-\frac{2}{4^{2k}}\right)t^{4k}\ .
\end{equation}
A closed-form expression for $S_1(t)$ can be found in Appendix \ref{identities}, see \eqref{S1-appendix}. In terms of this function, the first line in $\mathcal{\hat B}_{2m}(t^2)$ is
\begin{equation}
     B_1 = -2c_{3/2}(-4)^{-m-1}\sum_{n=0}^\infty \frac{(-4)^{-n}}{\Gamma(4n + 3)}\partial^{4n+4m+1}_t(t^{4n+4m+1}S_1(t))\ .
\end{equation}
The second line in \eqref{hat-B2m} can be written similarly as
\begin{equation}
     B_2 = -2c_{3/2}(-4)^{-m}\sum_{n=0}^\infty \frac{(-4)^{-n}}{\Gamma(4n + 1)}\partial^{4n+4m-1}_t(t^{4n+4m-1}S_2(t))\ .
\end{equation}
For more details, see Appendix \ref{identities}. In particular, the function $S_2(t)$ is given in \eqref{S2-appendix}. By applying the Laplace transform, integrating by parts and summing over $n$, we obtain the solution
\begin{align}
    \hat a_{2m}& = \int^\infty_0 e^{-t}\mathcal{\hat B}_{2m}(t^2)dt \\ 
    =& \frac{-2 c_{3/2}}{(-4)^m}\int^\infty_0\sum_{n=0}^\infty e^{-t} \left(\frac{(-4)^{-n-1}}{\Gamma(4n + 3)}t^{4n+4m+1}S_1(t) - \frac{(-4)^{-n}}{\Gamma(4n + 1)}t^{4n+4m-1}S_2(t)\right)dt \nonumber \\ \nonumber
    =& \frac{-c_{3/2}}{(-4)^{m+1}}\int^\infty_0 e^{-t}t^{4 m} K_1(t)dt = c_{3/2} \Bigg(\frac{(4m)!}{(-4)^{m+1}} -\frac12 \pi^{4m+1} \text{Li}_{-4m}(-e^\pi) \Bigg) \ .
\end{align}
Here, we have denoted
\begin{equation}
    K_1(t)= \frac{\sinh t - \sin t}{\cosh t- \cos t} \ .
\end{equation}
To do the final integral, we made use of the identity \eqref{integral-identity-1} and further simplified the expressions. The answer for the double-trace coefficients (recalling \eqref{notations}) reads 
\begin{equation}\label{32-double-traces-sum-rules-even}
\Delta_\phi=\frac32\, : \quad \hat a_{[\phi\phi]_{2\Delta_\phi+4m}} 
= \frac{192\sqrt{\pi}}{175} \Bigg( (-4)^m + \frac{(2\pi)^{4m+1}}{(4m)!}\text{Li}_{-4m}(-e^\pi)\Bigg) \ .
\end{equation}
Remarkably, we recover precisely the coefficients of the asymptotic model. In more detail, recall that $\hat F_k$ in the discussion above correspond to the asymptotic stress-tensor contribution to the sum rules. These are to be added to the identity contribution, which yields the GFF OPE coefficients. In accord with this, we find
\begin{equation}
    \Delta_\phi=\frac32\, : \quad \hat a_{[\phi\phi]_{4m+3}} = a_{[\phi\phi]_{4m+3}}^{\text{AM}} -  a_{[\phi\phi]_{4m+3}}^{\text{GFF}}\,,
\end{equation}
where $a_{[\phi\phi]_{4m+3}}^{\text{AM}}$ and $a_{[\phi\phi]_{4m+3}}^{\text{GFF}}$ are given by \eqref{AM-double-trace-coefficients-32} and \eqref{GFF-one-pt-coefficients}, respectively.

\subsection*{Odd $\bf{\mathcal{\hat B}_{2m+1}}$}

The procedure is very similar as for the even coefficients $\mathcal{\hat B}_{2m}$, so we will be brief. In this case, the sum \eqref{Borel-transform-holography} becomes
\begin{align}
     \nonumber \mathcal{\hat B}_{2m+1}(t^2) &= \sum_{k=0}^{\infty} \frac{\zeta(4k)}{\pi^{4k}} \left(2-4^{1-2k}\right)\frac{\hat F_{2k+2m}}{\Gamma(4k+1)}t^{4k} \\ &- \sum_{k=1}^{\infty}  \frac{\zeta(4k-2)}{\pi^{4k-2}} \left(2-4^{2-2k}\right)\frac{\hat F_{2k+2m-1}}{\Gamma(4k-1)}t^{4k-2}\ .
\end{align}
Following the same steps as above, the solution for the remaining double-trace coefficients can be expressed as
\begin{equation}
    \hat a_{2m+1}  =  \int^\infty_0 e^{-t}\mathcal{\hat B}_{2m+1}(t^2)dt = - \frac{1}{2}(-4)^{-m-1}\int^\infty_0 e^{-t}t^{4 m+2}K_2(t)dt\,,
\end{equation}
where
\begin{equation}
    K_2(t) = \frac{\sinh t + \sin t}{\cosh t- \cos t}\ .
\end{equation}
\newline
Evaluating the integral by means of \eqref{integral-identity-2} and plugging back into the relation \eqref{notations}, we obtain
\begin{equation}\label{32-double-traces-sum-rules-odd}
\Delta_\phi=\frac32\, : \quad \hat a_{[\phi\phi]_{2\Delta_\phi+4m+2}} 
= -\frac{96\sqrt{\pi}}{175} \Bigg( (-4)^{m+1} + \frac{2(2\pi) ^{4 m+3}}{{(4 m+2)!}} \text{Li}_{-4 m-2}\left(-e^{\pi }\right) \Bigg)\ .
\end{equation}
Again, this exactly reproduces the asymptotic model, i.e.
\begin{equation}
   \Delta_\phi=\frac32\, : \quad \hat a_{[\phi\phi]_{4m+5}} = a_{[\phi\phi]_{4m+5}}^{\text{AM}} -  a_{[\phi\phi]_{4m+5}}^{\text{GFF}}\,,
\end{equation}
with $a_{[\phi\phi]_{4m+5}}^{\text{AM}}$ and $a_{[\phi\phi]_{4m+5}}^{\text{GFF}}$ given in \eqref{AM-double-trace-coefficients-32} and \eqref{GFF-one-pt-coefficients}.

\subsection*{Solutions for other external scaling dimensions}

Having shown how the asymptotic holographic sum rules are solved in the case of conformal dimension $\Delta_\phi=3/2$, let us briefly discuss other values of $\Delta_\phi$. In Appendix \ref{Holographic solutions} we show how to compute the double-trace coefficients $a_{[\phi\phi]_{2\Delta_\phi+2m}}$ when the stress-tensor coefficients are given by \eqref{asymptotic-stress-tensor-coefficients} with $\Delta_\phi=5/2$. These again reproduce the asymptotic model of the same $\Delta_\phi$, that is detailed in Appendix \ref{sec:asymptotic-model-52}.
\smallskip

For general half-integer dimensions, the sum rules may again be solved explicitly. In the case $\Delta_\phi=2r-\frac12,\ r>1$, we obtain the solution\footnote{Note that $a_{[\phi\phi]_{2\Delta_\phi}}$ is not fixed by the KMS condition.}
\begin{align}
    \hat a_{2m} &= -c_{\Delta_\phi}(-4)^{-m-5r+4}\sum_{i=0}^{4r-4} s_{i,r} I^{(1)}_{4m+i} + \sum_{i=1}^{r-1} \frac{\Lambda^0_i}{2^{4i}} \  a^{\text{GFF}}_{\Delta_\phi-2i,\,2m}\,,\label{general-am-eqn-1} \\
    \hat a_{2m+1} &= -\frac{c_{\Delta_\phi}}{2}(-4)^{-m-5r+4}\sum_{i=0}^{4r-4} s_{i,r} I^{(2)}_{4m+2+i}+ \sum_{i=1}^{r-1} \frac{\Lambda^0_i}{2^{4i}} \  a^{\text{GFF}}_{\Delta_\phi-2i,\,2m+1}\,,\label{general-am-eqn-2}
\end{align}
while for the half-integer dimensions of the form $\Delta_\phi=2r+\frac12,\ r>0$, the solution reads
\begin{align}
    \hat a_{2m} &= -\frac{c_{\Delta_\phi}}{2}(-4)^{-m-5r+2}\sum_{i=0}^{4r-2} s_{i,r + \frac{1}{2}} I^{(2)}_{4m+i} + \sum_{i=1}^{r} \frac{\Lambda^0_i}{2^{4i}} \  a^{\text{GFF}}_{\Delta_\phi-2i,\,2m}\,, \label{general-am-eqn-3}\\ 
    \hat a_{2m+1} &= -c_{\Delta_\phi}(-4)^{-m-5r+1}\sum_{i=0}^{4r-2} s_{i,r+\frac12} I^{(1)}_{4m+2+i}+ \sum_{i=1}^{r} \frac{\Lambda^0_i}{2^{4i}} \  a^{\text{GFF}}_{\Delta_\phi-2i,\,2m+1}\ .\label{general-am-eqn-4}
\end{align}
Here, the $s_{i,r}$ denote the coefficients of the polynomial defined by the expression
\begin{equation}
    (t \partial_t)^{4r-4}\frac{e^{-t}}{t^{4r-2}} = \frac{e^{-t}}{t^{4r-2}}\sum_{i=0}^{4r-4}s_{i,r} t^i\ .
\end{equation}
They may be extracted explicitly via
\begin{equation}
    s_{i,r} = \left.\frac{1}{i!}\partial_t^i\left( e^t t^{4r-2}(t \partial_t)^{4r-4}\frac{e^{-t}}{t^{4r-2}}\right)\right|_{t=0}\ .
\end{equation}
On the other hand, $I^{(1)}_n$ and $I^{(2)}_n$ are defined in \eqref{integral-identity-1} and \eqref{integral-identity-2}. Let us note that \eqref{general-am-eqn-1}–\eqref{general-am-eqn-4} typically consist of two types of terms, the first that involves $I^{(i)}_n$ and the second involving the coefficients $a^{\text{GFF}}_{\Delta_\phi-2i,\,m}$. In the example $\Delta_\phi = 3/2$ that we detailed above, only the first kind of terms is present. For generic half-integer $\Delta_\phi$, special care needs to be taken when gamma functions in the sum rules \eqref{hol_sum_rules} exhibit singularities (as we have mentioned, the poles cancel between the numerator and the denominator in \eqref{hol_sum_rules}). The proper treatment of these terms in the sum rules gives rise to the second type of contributions in \eqref{general-am-eqn-1}–\eqref{general-am-eqn-4}. To see this explicitly in the simplest nontrivial case, we refer to the example with $\Delta_\phi = 5/2$ discussed in Appendix \ref{Holographic solutions}.
\smallskip

Formulas \eqref{32-double-traces-sum-rules-even} and \eqref{32-double-traces-sum-rules-odd}, as well as their generalizations \eqref{general-am-eqn-1}-\eqref{general-am-eqn-4} are some of the main results of the present section. Notice that we have found that the particular solution \eqref{regularized_solution}, without adding to it any homogeneous piece \eqref{homogeneous-solutions}, reproduces the asymptotic model of Section \ref{S:Toy models of holography}. The same was true in the GFF example, where \eqref{regularized_solution} gives the exact GFF double-trace coefficients. From these facts, it is natural to expect that in any theory, the particular solution \eqref{regularized_solution} to the KMS sum rules is the one among infinitely many solutions that gives the physical OPE coefficients.

\subsection{Application to the refined model}
\label{SS:Application to the refined model}

In this subsection, we briefly discuss the question: ‘What happens to solutions to sum rules \eqref{hol_sum_rules} if one adjusts a finite number $N$ of stress-tensor OPE coefficients?' \smallskip

By linearity, we may take the solution to the asymptotic sum rules obtained above and add to it a solution to the sum rules with $F_k$ given by a finite version of the sum \eqref{stress rhs}, with the replacement $\Lambda_n\to\delta\Lambda_n = \Lambda_n - \Lambda_n^0$ for $ n=1,\ldots,N$ and $\Lambda_n\to0$ for $n>N$. To emphasize their nature as a correction, let us relabel the resulting coefficients as $F_k \to \delta F_k$. Thus, we consider the sum rules with
\begin{equation}\label{finite-sum-rule-corrections}
    \delta F_k = -\sum_{n=1}^N\frac{\delta \Lambda_n}{2^{4n}}\frac{\Gamma({4n} - 2\Delta_\phi+1)}{\Gamma({4n} - 2k - 2\Delta_\phi)} \equiv \sum_{n=1}^N \delta F_k^{(n)} \ .
\end{equation}
Again by linearity, the sum rules can be solved for each $n$ individually. In the case where the scaling dimension of the field is given by $\Delta_\phi = p + \frac12$, with $p \in \mathbb{N}$, we distinguish between two regimes: (i) $2n < \Delta_\phi$, and (ii) $2n > \Delta_\phi$. In the first regime, both the numerator and the denominator in \eqref{finite-sum-rule-corrections} are singular, but we may recast $\delta F_k^{(n)}$ in the form which is well-defined
\begin{equation}\label{infinite-sum-rule-corrections}
   2n < \Delta_\phi\, : \quad \delta F_k^{(n)} = \frac{\delta \Lambda_n}{2^{4n}}\frac{\Gamma(2k - 4n + 2\Delta_\phi + 1)}{\Gamma(2\Delta_\phi - 4n )}\ .
\end{equation}
Up to an overall constant, the resulting sum rules coincide with those of the GFF model with conformal dimension $\Delta_\phi - 2n$. This term contributes to the double-trace coefficients as
\begin{equation}\label{Borel_corrections-1}
   2n < \Delta_\phi\, : \quad  \delta a_m^{(n)} =  2^{-2m-2\Delta_\phi+1} \delta \Lambda_n\,\zeta (2m + 2\Delta_\phi - 4n) (2\Delta_\phi-4n)_{2m}\ .
\end{equation} 
For the second regime $2n > \Delta_\phi$, we have $\delta F_k^{(n)} = 0$ for $k>2n-p-1$, owing to the singularity of the gamma function appearing in the denominator in \eqref{finite-sum-rule-corrections}. Resummation methods are not required for the remaining finite system, as the solution given by \eqref{particular_solution} is already convergent.  Thus, in this regime the corrections to the coefficients are given by
\begin{align}\label{Borel_corrections-2}
    \nonumber 2n>\Delta_\phi\, : \quad \delta a_m^{(n)} &= \sum_{k=0}^{2n-m-p} (-1)^k \frac{\zeta(2k)}{\pi^{2k}} \left(2-4^{1-k}\right)\delta F_{k+m-1}^{(n)}\,,\quad m\leq 2n-p\,,
    \\ \delta a_m^{(n)} &= 0\,, \quad m > 2n-p\ .
\end{align}
Putting \eqref{Borel_corrections-1} and \eqref{Borel_corrections-2} together, the solution to the sum rules \eqref{finite-sum-rule-corrections} takes the form
\begin{align}
    \nonumber \delta a_m = &\sum_{n=1}^{\lfloor\Delta_\phi/2 \rfloor}2^{-2m-2\Delta_\phi+1} \delta \Lambda_n\,\zeta (2m + 2\Delta_\phi - 4n) (2\Delta_\phi-4n)_{2m}\ \\
     + &\sum_{n=\lfloor\Delta_\phi/2\rfloor+1}^{N}\sum_{k=0}^{2n-m-p} (-1)^k \frac{\zeta(2k)}{\pi^{2k}} \left(2-4^{1-k}\right)\delta F_{k+m-1}^{(n)}\,, \label{Borel_corrections}
\end{align}
where, in the second line, whenever $2n-m-p<0$, the sum is simply set to zero. If we express this result in terms of the original double-trace coefficients $a_{[\phi\phi]_{2\Delta_\phi+2m}}$, using the relation \eqref{notations}, and subsequently sum over them with the corresponding powers $\tau^{2m}$, we recover the two-point function presented in equation \eqref{low-operator-corrected-CAM}. This result is identical to the refined asymptotic model.\footnote{For a detailed discussion of the analytic properties of the resulting two-point function, see Section \ref{SS:Refinements of the model}.} Intuitively, in the limit $N \gg 1$, where a large number of stress-tensor coefficients receive corrections, one might expect the double-trace coefficients to approach their true holographic values. However, this expectation is not fulfilled. Let us examine this more closely. First, note that for fixed \( n > \lfloor \Delta_\phi / 2 \rfloor \), the function \( \delta F_k^{(n)} \) exhibits factorial growth of the form \( (2k)! \), which arises from the presence of \( \Gamma(4n - 2k - 2\Delta_\phi) \) in its denominator (see \eqref{finite-sum-rule-corrections}). As a result, within the sum over $k$ in \eqref{Borel_corrections}, the dominant contribution arises from the term with the largest value of $k = 2n-m-p$, and
 \begin{equation}
    \left.\delta F_{k+m-1}^{(n)}\right|_{k=2n-m-p} = \delta F^{(n)}_{2n-p-1} = - \frac{\delta \Lambda_n}{2^{4n}}\Gamma(4n - 2\Delta_\phi +1)\,, \quad 2n > \Delta_\phi\ .
 \end{equation}
On the other hand, the difference $|\delta \Lambda_n| = |\Lambda_n - \Lambda_n^0|$ is expected to grow exponentially as $ \sim 4^n$.
In turn, this implies that when $\delta a_m$ is considered as a function of $N\gg1$, the sum in the first line of \eqref{Borel_corrections} can be treated as a constant, while the dominant contribution in the sum in the second line arises from the term with the largest $n$, that is, 
\begin{equation}\label{dominant-correction-am}
    \delta a_m \sim \frac{\delta\Lambda_N}{(2\pi)^{4N}}\Gamma(4N-2\Delta_\phi+1) \sim (4N)!\ .
\end{equation}
We conclude that the asymptotic model refined in the above way is not well-defined in the limit $N\to\infty$ as the coefficients $\delta a_m$ do not converge to a finite limiting value. Furthermore, adjusting a finite number of stress-tensor coefficients and subsequently applying Borel regularization to \eqref{Borel_corrections} will not lead to a meaningful result in the limit $N\to\infty$, as
\begin{equation} 
\lim_{N\to\infty}\int_0^{\infty}e^{-t}\sum_{n=1}^N \frac{\delta a_m^{(n)}}{(4n)!}t^{4n} = \lim_{N\to\infty}\sum_{n=1}^N\int_0^{\infty}e^{-t} \frac{\delta a_m^{(n)}}{(4n)!}t^{4n} = \sum_{n=1}^N \delta a_m^{(n)}\,,
\end{equation}
diverges. Hence, a more promising way to approach the actual holographic correlator is to consider subleading large $n$ corrections to the asymptotic model, as discussed at the end of Section \ref{S:Toy models of holography}.

\section{Discussion}
\label{S:Discussion}

In this work, we have constructed models of holographic thermal two-point functions that capture the asymptotic behavior of OPE coefficients of composites of the stress tensor. As an input, we took the large $n$ asymptotics of thermal one-point coefficients of operators of the schematic form $[T_{\mu\nu}]^n$, determined in \cite{Ceplak:2024bja}, and as an output obtained one-point coefficients of double trace-operators $[\phi\phi]_{n,\ell}$  which constitute the rest of the $\phi\times\phi$ OPE at large $C_T$. We have explored two methods, both of which rely on the KMS invariance of the two-point function. The first approach involves summing over thermal images of the stress-tensor sector and subsequent cancellation of the bouncing singularities by double-trace terms.
The second method involves solving the KMS sum rules directly. This requires Borel resummation in order to regularize divergent formal solutions. Both approaches produce the same results.\footnote{Holographic computation of the correlator, e.g. via solving the relevant PDE (see Appendix \ref{A:Laplace equation on the AdS-Schwarzschild black hole}) requires regularity in the bulk with the period of the angular variable/Euclidean time $\tau$ set to $\beta$. This produces the unique correlator which has the stress-tensor sector consistent with the near-boundary OPE and is periodic in $\tau$. Our paper develops an algebraic way of completing the stress-tensor sector to the full KMS-periodic correlator.}
\smallskip

In Section \ref{SS:Properties of two-point functions}, we listed expected properties of thermal two-point functions, all of which are satisfied by our asymptotic model.\footnote{As discussed in Section \ref{S:Toy models of holography}, the asymptotic model satisfies the weak form of boundedness, namely $|g_\beta(\tau)|<\infty$ as Im$\,\tau\to\infty$, and not necessarily \eqref{Regge-bound}.} Let us note that that the model does not satisfy the property discussed in \cite{Marchetto:2023xap} in the context of thermal CFT Tauberian theorems. Namely, one-point coefficients of the model do not become sign-definite at asymptotically large $\Delta$. This is clear from our input data, as already the multi-stress-tensor coefficients are asymptotically sign-alternating. We conclude that sign-definiteness does not follow from properties listed in Section \ref{SS:Properties of two-point functions}.
\smallskip

Our results open several interesting avenues for further  investigation. One of the main open questions  is to move from the asymptotic model studied in this work to real holography. As was stressed several times in this paper, the models that we discussed do not produce exact holographic one-point coefficients because the multi-stress-tensor data used for their construction  only captures the asymptotic behavior of the $[T_{\mu\nu}]^n$ data at large $n$. 
Therefore, one would like  to systematically improve the models by inputting more information about the stress-tensor sector. The discussion of Sections \ref{SS:Refinements of the model} and \ref{SS:Application to the refined model} suggests that the appropriate way of incorporating this data is  via an asymptotic large-$n$ expansion such as
\begin{equation}\label{discussion-large-n-asymptotics}
    \Lambda_n \simeq \Lambda_n^0 \sum_{s=0}^\infty \frac{\alpha_s}{n^s}\ .
\end{equation}
To this end, the first step is to extract from the exact stress-tensor data the asymptotic expansion coefficients $\alpha_s$ appearing in \eqref{discussion-large-n-asymptotics}. Let us note that, in principle, there may be additional subleading terms in \eqref{discussion-large-n-asymptotics} with slower exponential growth. While these terms can provide valuable insights into subleading poles of the correlator, they are particularly challenging to identify due to the lack of data for a more accurate approximation. Once the refined stress-tensor asymptotics \eqref{discussion-large-n-asymptotics} is obtained, one can construct the corresponding asymptotic model following the methodology outlined in Section \ref{SS:Refinements of the model} or \ref{SS:Application to the refined model}, and  extract the corresponding double-trace coefficients. 
We leave this for future work.
\smallskip

An alternative approach to determining the holographic one-point coefficients of double-trace operators involves numerically solving the wave equation on the AdS-Schwarzschild black hole background. We discuss this method, following \cite{Parisini:2023nbd}, in Appendix \ref{A:Laplace equation on the AdS-Schwarzschild black hole}. A meaningful comparison with the present work requires both the refinements described in \eqref{discussion-large-n-asymptotics} and improved numerical precision on the wave equation side. A preliminary comparison of the double-trace coefficients shows promising agreement; see Appendix \ref{A:Laplace equation on the AdS-Schwarzschild black hole}.
\smallskip

There are various generalizations one may wish to consider. The models in the main text were restricted to have half-integer scaling dimension $\Delta_\phi=p+1/2$ of the external field, in which case all the computations may be carried out explicitly. For simplicity, we often displayed results for the special case $\Delta_\phi=3/2$. For other half-integer scaling dimensions both the asymptotic model and the solution to the KMS sum rules can be obtained along the same lines. For the case $\Delta_\phi = 5/2$, the two computations are detailed in appendices \ref{sec:asymptotic-model-52} and \ref{Holographic solutions} (and again produce the same two-point function).  We believe that explicit expressions for the two-point function and double-trace coefficients for generic $p$ are not difficult to obtain. It is an interesting challenge to find solutions for generic real values of $\Delta_\phi$, in which case the stress-tensor sector develops a branch cut. In another direction, changing the spacetime dimension introduces minimal modifications in our discussion, as long as the asymptotic stress-tensor data is available. For instance, the asymptotic model in $d=6$ is described in Appendix \ref{A:Models in six dimensions}. 
\smallskip

Another extension  is to analyze the two-point function at non-vanishing spatial separation $\vec{x}$, thus probing the double-trace coefficients of operators of different spin (as opposed to spin-averaged coefficients considered in this work).
One may wonder if the corresponding generalization of the asymptotic model will exhibit quasinormal modes similar to the ones of  holographic systems.
\smallskip

Going beyond holography, large parts of our discussion apply to general CFTs at finite temperature. On the one hand, the discussion of Section \ref{SS:Sum over images} suggests a set of symmetry, analyticity and spectral properties that might uniquely specify a thermal two-point function, as well as the construction of the latter via a modification of the method of images. On the other, the approach of Section \ref{S:Sum rules} demonstrates that KMS sum rules can in some cases be effectively solved by methods of linear algebra. Moreover, at least in the models we studied, the obtained solution automatically has good analytic structure. It is an interesting problem to try and identify other theories that can be approached by these techniques.
\smallskip

A major ingredient of this paper has been Borel summation of the divergent series to obtain the KMS-invariant form of the correlator. Recently, similar techniques have been used in the context of boost-invariant flows 
\cite{Heller:2013fn,Heller:2015dha,Basar:2015ava,Aniceto:2015mto,Florkowski:2016zsi,Spalinski:2017mel,Casalderrey-Solana:2017zyh,Heller:2018qvh}. It would be interesting to explore potential connections between our approach and those developments.

\section*{Acknowledgements}

We wish to thank N. \v Ceplak, C. Esper, A. Grinenko, M. Kulaxizi, H. Liu, V. Niarchos, C. Papageorgakis, E. Parisini, F. Russo, S. Solodukhin, S. Valach, B. Withers for discussions, correspondence and comments on the manuscript. 
This publication has emanated from research conducted with the financial support of Taighde Éireann – Research Ireland under Grant number SFI-22/FFP-P/11444.

\appendix

\section{Asymptotic Models for Various \texorpdfstring{$\Delta_\phi$}{Delta\_phi} and \texorpdfstring{$d$}{d}}
\label{sec:asymptotic-models-delta-d}

\subsection{Asymptotic Model for \texorpdfstring{$\Delta_\phi = \frac{5}{2}$}{Delta\_phi = 5/2}}
\label{sec:asymptotic-model-52}

In this appendix, we give details about the asymptotic model in four dimensions with $\Delta_\phi=5/2$. By summing the asymptotic stress-tensor coefficients \eqref{asymptotic-stress-tensor-coefficients}, we obtain the stress-tensor sector
\begin{equation}\label{stress-tensor-sector-52}
    g_T(\tau) = -\frac{8192\sqrt{\pi}\alpha_0}{693} \frac{1-4 \tau^4}{\tau  \left(1+4 \tau ^4\right)^3}\ .
\end{equation}
Next, we perform the sum over thermal images of \eqref{stress-tensor-sector-52}. To write the result, we make use of the function
\begin{small}
\begin{align*}
    f(\tau) & = \frac{128 \sqrt{\pi}\alpha_0}{693} \Bigg(64\psi(\tau) - 32\psi\left(\tau-\frac{1+i}{2}\right)- 32\psi\left(\tau -\frac{1-i}{2}\right) - 7i \psi^{(1)} \left(\tau -\frac{1+i}{2}\right)\\ 
    & + 7i \psi^{(1)}\left(\tau -\frac{1-i}{2}\right) - \frac{256 \tau ^5-584 \tau ^4+712 \tau ^3-488 \tau ^2+196 \tau -42}{(2\tau^2 - 2\tau +1)^3}\Bigg)\,,
\end{align*}
\end{small}\newline
where $\psi^{(k)}(z)$ denotes the polygamma function. In terms of the function $f(\tau)$, the sum of images over \eqref{stress-tensor-sector-52}, with the added contribution from the identity, can be written as
\begin{equation}\label{image-sum-52}
    g_\beta^{\text{images}}(\tau) = g^{\text{GFF}}_{\Delta_\phi=5/2}(\tau) + f(\tau) + f(1-\tau)\ .
\end{equation}
The sum over images has third order poles at $(1\pm i)/2$. These are canceled by means of the subtraction \eqref{gbeta-definition}. In the case at hand, the subtraction term can be written as a linear combination of $h$, $h^2$ and $h^3$, where $h(\tau)$ was given in \eqref{pole-sum-of-images}. Concretely
\begin{align}\label{AM-4d-52}
    g_\beta(\tau) &= g_\beta^{\text{images}}(\tau)\\
    & + \frac{512\pi^{5/2}\alpha_0}{693} \Bigg(4\pi h(\tau)^3 + (6\pi\coth\pi - 7)h(\tau)^2 + (2\pi- 7\coth\pi)h(\tau) \Bigg)\ . \nonumber
\end{align}
This is the asymptotic model for $\Delta_\phi=5/2$. One shows that the only singularities of \eqref{AM-4d-52} on the vertical strip are the KMS poles at $\tau=0,1$. The double-trace one-point coefficients of \eqref{AM-4d-52} can readily be extracted - they are written in \eqref{52-double-trace-coefficients}.

\subsection{Models in six dimensions}
\label{A:Models in six dimensions}

In this appendix we construct the asymptotic model in six dimensions. The main difference between four and six dimensions is in the behavior of multi-stress-tensor one-point coefficients. In $d=6$, the asymptotic form of the latter reads
\begin{equation}\label{6d-asymptotics}
    \Lambda_n^0 = \tilde{c}_{\Delta_\phi} n^{2\Delta_\phi-4}\ .
\end{equation}
Although the function $\tilde{c}_{\Delta_\phi}$ is not explicitly known, its precise form is not essential for the purposes of our analysis. We shall fix $\Delta_\phi=\frac{5}{2}$ (recall that the unitarity bound for scalars in $d=6$ is $\Delta_\phi=2$). Other values of $\Delta_\phi$ will be briefly discussed at the end.
\smallskip

Asymptotic coefficients \eqref{6d-asymptotics} give the stress-tensor sector of the two-point function
\begin{equation}\label{6d-gT}
    g_T(\tau) = \tau^{-2\Delta_\phi} \sum_{n=1}^\infty \Lambda_n^0 \tau^{6n} = \tilde{c}_{\frac52} \frac{\tau}{\left(1-\tau^6\right)^2}\ .
\end{equation}
To proceed, we add the identity exchange contribution and perform the sum over thermal images
\begin{equation}\label{6d-image-sum}
    g_\beta^{\text{images}}(\tau) = \sum_{m=-\infty}^\infty \left( \frac{1}{|\tau+m|^5} + \tilde{c}_{\frac52} \frac{|\tau+m|}{\left(1-(\tau+m)^6\right)^2}\right)\ .
\end{equation}
The sum over the first term gives the GFF two-point function. The sum over the second can also be performed exactly in terms of polygamma functions. To write the final result, let $\omega=e^{\frac{i\pi}{3}}$. Then the image sum \eqref{6d-image-sum} gives
\begin{align}\label{6d-image-sum-result}
   g_\beta^{\text{images}}&(\tau) = g^{\text{GFF}}_{\Delta_\phi=5/2}(\tau) + \frac{\tilde{c}_{\frac52}}{18}\Bigg(\frac{1}{\tau^2} + \frac{1}{(\tau-1)^2} + \frac{3}{\left((\tau-\omega)(\tau-\omega^{-1})\right)^2}\nonumber\\
    & + \frac{4}{(\tau-\omega)(\tau-\omega^{-1})} + \frac{2}{\tau-\omega^2} + \frac{2}{\tau-\omega^{-2}} - \frac{4}{\tau+1} + 4\psi(\tau-1) + 4\psi(-\tau-1)\nonumber\\
    & - 2\psi(\tau-\omega) - 2\psi(-\tau-\omega) - 2\psi\left(\tau+\omega^2\right) - 2\psi\left(-\tau+\omega^2\right) \Bigg)\ .
\end{align}
The difference compared to the image sum in four dimensions, arising due to the different form of asymptotics \eqref{6d-asymptotics} is in the pole structure of \eqref{6d-image-sum-result}. This function has poles at sixth roots of unity. In addition, the pole at $\tau=0$ does not correspond to the holographic spectrum - it contains a non-physical operator of dimension 3. All these singularities are canceled by the procedure of Section \ref{SS:Sum over images}. The final result for the asymptotic model reads
\begin{equation}\label{6d-AM}
    g_\beta(\tau) = g_\beta^{\text{images}}(\tau) - \tilde{c}_{\frac52}\left(\frac{1}{18} g^{\text{GFF}}_{\Delta=1}(\tau) - 2\gamma \left(H(\tau) - \frac12\right)  + \frac{8\pi^2}{9} \left(H(\tau)^2 - \frac14\right)\right)\ .
\end{equation}
The subtraction term proportional to $g^{\text{GFF}}_{\Delta=1}(\tau)$ arises as the sum of images of the subleading pole of $g_\beta^{\text{images}}(\tau)$ at the origin. The other two terms make use of the function
\begin{equation}\label{6d-pole-cancelling-function}
    H(\tau) = \frac{1+ 2 e^{\pi  \left(\sqrt{3}+2i\tau\right)}+e^{4i\pi\tau}}{2\left(e^{-\pi\sqrt{3}} + e^{2i\pi\tau}\right) \left(e^{\pi\sqrt{3}}+e^{2i\pi\tau}\right)}\,, \quad \gamma = \pi\coth(\pi\sqrt{2}) + 2\pi - \sqrt{3}\,,
\end{equation}
which is KMS-invariant and has first order poles at $\omega^{\pm1}$. They arise as image sums of singular pieces of $g_\beta^{\text{images}}(\tau)$ at $\omega^{\pm1}$. The asymptotic model \eqref{6d-AM} satisfies all conditions of Section \ref{SS:Properties of two-point functions} and has the stress-tensor content given in \eqref{6d-asymptotics}. In particular, one can verify the boundedness property \eqref{Regge-bound}.
\smallskip

We have tested the above construction for other half-integer values of $\Delta_\phi=p+1/2$, where all the sums can be performed exactly. We find that the same pattern persists: the image sum analogous to \eqref{6d-image-sum} has singularities at $\omega^{\pm1}$, as well as subleading poles at $\tau=0$ that correspond to unphysical operators in the spectrum. The unphysical operators can be removed in a unique way by adding a linear combination of GFF two-point functions $g^{\text{GFF}}_{\Delta}$ with $\Delta=1,\dots,p-1$. On the other hand, the singularities at $\omega^{\pm1}$ are canceled uniquely by adding a linear combination of functions $\{H,H^2,\dots,H^{2(p-1)}\}$.

\section{Solutions to Sum Rules}
\label{Solutions to Sum Rules}

In this appendix, we solve the KMS sum rules for the asymptotic holographic model using Borel resummation. We explicitly work out the case $\Delta_\phi = \frac52$ from Section \ref{SS:Borel-toy-model}, deriving the corresponding thermal one-point coefficients. We also collect a set of identities and summation formulas used throughout the main text.

\subsection{Holographic solutions}\label{Holographic solutions}

The smallest representative of the second set of conformal dimensions given in \eqref{conformal_dimensions} is 
$\Delta_\phi = \frac52$. The multi-stress-tensor part of the sum rules is given by 
\begin{equation}
    \Delta_\phi = \frac52\, : \quad \widetilde{F}_k = -\sum_{n=1}^\infty \frac{\Lambda^0_n}{2^{4n}}\frac{\Gamma({4n} - 4)}{\Gamma({4n} - 2k - 5)}\,,
\end{equation}
where, once again, one can verify that all terms in the series are finite.
Since the system of equations is linear, the $n = 1$ term — in which both the numerator and denominator gamma functions exhibit singularities — can be treated separately and rewritten as follows:
\begin{equation}
    \widetilde{F}^{(1)}_k = \frac{\Lambda^0_1}{2^{4}}\Gamma(2k+2)\ .
\end{equation}
Its contribution to the double-trace coefficients can be determined independently. Note that this is proportional to the GFF result for $\Delta_\phi = \frac{1}{2}$ (see \eqref{GFF_coefs}).
It is once again useful to decompose the rest of the terms into two distinct parts
\begin{align}
    \nonumber\Delta_\phi = \frac52\, : \quad \widetilde{F}_{2k} &
    = -\frac{1}{2^{4k+8}}\sum_{n=0}^\infty \frac{\Lambda^0_{n+k+2}}{2^{4n}}\frac{\Gamma(4n+4k + 4)}{\Gamma(4n + 3)}\,,\\
    \widetilde{F}_{2k-1} &
    =-\frac{1}{2^{4k+4}}\sum_{n=0}^\infty \frac{\Lambda^0_{n+k+1}}{2^{4n}}\frac{\Gamma(4n+4k)}{\Gamma(4n + 1)}\ .
\end{align}
Our goal is to solve the model with the asymptotic multi-stress-tensor coefficients given in equation \eqref{asymptotic-stress-tensor-coefficients} and compare it to the asymptotic model. The procedure will be very similar to the $\Delta_\phi = \frac32$  case described in the main text. The key difference is that we now need to differentiate sums in order to extract the $n^2$ factor that appears in the asymptotic coefficients $\Lambda_n^0$. The double-traces take the form
\begin{equation}
    \widetilde{a}_{2m} = \int^\infty_0 e^{-t}\widetilde{\mathcal{B}}_{2m}(t^2)dt 
    =  -\frac{c_{\Delta_\phi}}{2}(-4)^{-m-1}\int^\infty_0 K_2(t) t^{4m+4} \left(t\partial_t\right)^2\left(\frac{e^{-t}}{t^4}\right)dt \,,
\end{equation}
and 
\begin{equation}
    \widetilde{a}_{2m+1} = \int^\infty_0e^{-t}\widetilde{\mathcal{B}}_{2m+1}(t^2)dt 
    = -c_{\Delta_\phi}(-4)^{-m-2}\int^\infty_0 K_1(t) t^{4m+6} \left(t\partial_t\right)^2\left(\frac{e^{-t}}{t^4}\right)dt \ .
\end{equation}
The integrals can be evaluated by applying the corresponding identities \eqref{integral-identity-1}, \eqref{integral-identity-2}, yielding 
\begin{align}
    \Delta_\phi = \frac52\, : \quad \widetilde{a}_{2m} &= -\frac{c_{\Delta_\phi}}{2}(-4)^{-m-3}\left(16I^{(2)}_{4m} + 7I^{(2)}_{4m+1} + I^{(2)}_{4m+2} \right)\,, \\ \nonumber
    \widetilde{a}_{2m+1} &= -c_{\Delta_\phi}(-4)^{-m-4}\left(16I^{(1)}_{4m+2} + 7I^{(1)}_{4m+3} + I^{(1)}_{4m+4} \right)\ .
\end{align}
Eventually, incorporating the $n=1$ term that was considered separately, the final result for the double-trace coefficients takes the form:
\begin{align}\label{52-double-trace-coefficients}
   \Delta_\phi = \frac52\, : \quad  a_{2m} &= \widetilde{a}_{2m} +\frac{\Lambda^0_1}{2^{4}}2^{-4m} \zeta (4m + 1) \Gamma (4m + 1)\,, \\ \nonumber
    a_{2m+1} &= \widetilde{a}_{2m+1} +\frac{\Lambda^0_1}{2^{4}}2^{-4m-1} \zeta (4m + 2) \Gamma (4m + 2)\ .
\end{align}
By re-expressing the result in terms of the original double-trace coefficients—using the definitions from \eqref{notations}—and summing them with the powers $\tau^{2m}$, we recover a two-point function that precisely reproduces the asymptotic model for $\Delta_\phi=5/2$ (see \eqref{AM-4d-52}).

\subsection{List of identities}\label{identities}

Here we collect the integrals and identities used in the derivations above. In Section \ref{SS:Borel-toy-model}, we consider the sum rules for the case $\Delta_\phi = \frac{3}{2}$. The multi-stress-tensor contributions $\hat{F}_k$ in \eqref{F2k} and \eqref{F2k1} can be written explicitly as
\begin{align}
    \nonumber & \frac{\hat{F}_{2k}}{c_{3/2}} = -\frac{1}{2^{4(k+1)}}\sum_{n=0}^\infty \frac{(-4)^{n+k+1}}{2^{4n}}\frac{\Gamma(4n+4k + 2)}{\Gamma(4n + 1)} =\frac{(-1)^k}{4\cdot 5^{2k+1}}\Gamma(4k+2) \cos((4k+2)\theta)\,,\\
    & \frac{\hat{F}_{2k-1}}{c_{3/2}} = -\frac{1}{2^{4(k+1)}}\sum_{n=0}^\infty \frac{(-4)^{n+k+1}}{2^{4n}}\frac{\Gamma(4n+4k + 2)}{\Gamma(4n + 3)} =\frac{(-1)^k}{4\cdot 5^{2k}}\Gamma(4k) \sin(4k\theta)\,,\label{Fk-appendix}
\end{align}
where $\theta = \text{arctan}(-1/3)$. The coefficient $c_{\Delta_\phi=3/2}$ was defined in \eqref{asymptotic-stress-tensor-coefficients}. In the solution of asymptotic holographic sum rules in Section \ref{SS:Borel-toy-model}, we make use of the following identities
\begin{align}
    \label{S1-appendix} & S_1(t) \equiv \sum_{k=0}^{\infty}\frac{\zeta(4k)}{(-4\pi^4)^{k}} \left(1-\frac{2}{4^{2k}}\right)t^{4k} \\
    & \hskip5cm =  - \frac{t}{2(1+i)} \frac{\sin\left(\frac{1+i}{2}t\right) + \sinh\left(\frac{1+i}{2}t\right)}{\cos t - \cosh t}\,,\nonumber \\ \label{S2-appendix} 
    & S_2(t) \equiv \pi^2\sum_{k=1}^{\infty}\frac{\zeta(4k-2)}{(-4\pi^4)^{k}} \left(1-\frac{2}{4^{2k-1}}\right)t^{4k-2} \\ 
    & \hskip5cm = \frac{1+i}{8}\frac{ t \left(\sin \left(\frac{1+i}{2} t\right)-\sinh\left(\frac{1+i}{2} t\right)\right)}{\cos t-\cosh t}\ . \nonumber
\end{align}
The one-point coefficients \eqref{32-double-traces-sum-rules-even} and \eqref{32-double-traces-sum-rules-odd} are obtained by means of the integrals
\begin{align}\label{integral-identity-1}
    & I^{(1)}_n \equiv \int_0^\infty e^{-t} t^n \frac{\sinh t - \sin t}{\cosh t - \cos t}\, dt\\
    & \hskip2cm = -i \left(-\frac{1+i}{2}\right)^n \psi^{(n)}\left(\frac{1+i}{2}\right) + i \left(-\frac{1-i}{2}\right)^n \psi^{(n)}\left(\frac{1-i}{2}\right) - n!\,,\nonumber\\[2pt]
    & I^{(2)}_n \equiv \int_0^\infty e^{-t} t^n \frac{\sinh t + \sin t}{\cosh t - \cos t}\, dt\label{integral-identity-2}\\
    & \hskip2cm = - \left(-\frac{1+i}{2}\right)^n \psi^{(n)}\left(\frac{1+i}{2}\right) - \left(-\frac{1-i}{2}\right)^n \psi^{(n)}\left(\frac{1-i}{2}\right) - n!\ .\nonumber
\end{align}
The same integrals appear in solutions \eqref{general-am-eqn-1}-\eqref{general-am-eqn-4} to KMS sum rules.

\section{Klein-Gordon equation on the planar AdS black hole}
\label{A:Laplace equation on the AdS-Schwarzschild black hole}

In Section \ref{SS:Multi-stress tensor OPE coefficients}, we reviewed some facts about OPE coefficients of the stress-tensor sector $g_T(\tau)$ of the thermal two-point function. These are obtained by solving the eigenvalue equation for the Laplacian on the planar AdS-Schwarzschild black hole in the near boundary expansion. In order to access coefficients of the double-trace sector, one needs to go beyond the boundary expansion. A numerical approach to this problem was developed in \cite{Parisini:2023nbd}. In this appendix, we briefly review the method before presenting and discussing the results it yields in the context of the models considered in the main text.

The metric of the five-dimensional Euclidean AdS black hole reads
\begin{equation}
    ds^2 = \frac{1}{z^2} \left(\frac{dz^2}{1-\frac{z^4}{z_H^4}} + \left(1 - \frac{z^4}{z_H^4}\right)d\tau^2 + \delta_{ab}dx^a dx^b\right)\,, \qquad a,b,=1,2,3\ .
\end{equation}
We denoted the coordinates by $(z,\tau,x^a)$. The constant $z_H$ is related to the temperature by $\beta = \pi z_H$ and the coordinate $\tau$ is periodic with the period $\beta$. To analyze the Laplacian eigenvalue problem, we set $r^2 = x^a x_a$ and introduce coordinates $(\rho,\varphi,R)$ related to $(z,\tau,r)$ by
\begin{equation}\label{cube-coordinates}
    z = 1 - \rho^2\,, \qquad \tau = \frac{\varphi}{2}\,, \qquad r = \frac{R}{1-R^2}\ .
\end{equation}
We look for solutions of the equation
\begin{equation}
    \Delta \Phi = \Delta_\phi (\Delta_\phi - 4) \Phi\,,
\end{equation}
which are spherically symmetric, $\Phi = \Phi(\rho,\varphi,R)$, are regular in the bulk, and behave as a delta distribution on the boundary $\partial=\{z=0\}$,
\begin{equation}
    \Phi|_\partial = \tilde G_{\text{AdS}}|_\partial\,, \qquad \tilde G_{\text{AdS}} \equiv \frac{z^{\Delta_\phi}}{\left(z^2 + f(\varphi) + r^2\right)^{\Delta_\phi}}\ .
\end{equation}
The function $\tilde G_{\text{AdS}}$ is obtained from the AdS propagator
\begin{equation}
     G_{\text{AdS}} \equiv \frac{z^{\Delta_\phi}}{\left(z^2 + \frac{\varphi^2}{4} + r^2\right)^{\Delta_\phi}}\,,
\end{equation}
by replacing the term $\varphi^2/4$ in the denominator by a $2\pi$-periodic function $f(\varphi)$ with the same small-$\varphi$ behavior. The coordinates \eqref{cube-coordinates} take values in the parallelepiped $\rho,R\in[0,1]$, $\varphi\in[0,2\pi]$, with one pair of opposite faces being identified (corresponding to the periodic coordinate $\varphi$). This makes the Klein-Gordon equation suitable for numerical study, once the boundary conditions on the faces are determined.
\smallskip

In order to discuss the boundary conditions, we start from the fact that, since the planar black hole is an asymptotically AdS space, every solution to the wave equation behaves near the boundary as
\begin{equation}\label{asymptotic-behaviour-phi}
    \Phi \sim z^{\Delta_\phi} \Phi_+(\tau,\vec{x}) + z^{4-\Delta_\phi} \Phi_-(\tau,\vec{x})\ .
\end{equation}
We put $\Phi = \Psi + \tilde G_{AdS}$ and solve the equation for $\Psi$ instead of $\Phi$. Let us assume that $\Psi$ splits in the same way as $\Phi$
\begin{equation}\label{asymptotic-behaviour-psi}
    \Psi \sim z^{\Delta_\phi} \Psi_+(\tau,\vec{x}) + z^{4-\Delta_\phi} \Psi_-(\tau,\vec{x})\ .
\end{equation}
Depending on whether $\Delta_\phi$ is smaller or larger than two, different terms in the expression \eqref{asymptotic-behaviour-psi} are leading as $z\to0$. We confine ourselves to the case $\Delta_\phi=3/2$ and put
\begin{equation}
    \Psi = z^{\Delta_\phi} H\ .
\end{equation}
We look for a solution with $\Psi_-=0$. Together with regularity in the bulk, this implies the following set of Dirichlet and Neumann boundary conditions on $H$
\begin{equation}\label{Dirichlet-and-Neumann}
    H|_{R=1} = \partial_R H |_{R=0} =\partial_\rho H |_{\rho=0} = \partial_\rho H |_{\rho=1} = 0\ .
\end{equation}
The differential equation for $H$ with boundary conditions \eqref{Dirichlet-and-Neumann} can be solved numerically. One possibility is to discretise the equation on a grid consisting of $N^3$ points. We use the uniform distribution of points in $\varphi$- and $R$-directions and the Chebyshev collocation in the $\rho$-direction. Finally, the thermal two-point function is related to the solution by
\begin{equation}\label{H-gbeta-relation}
    g_\beta(\tau,r) = H(\rho=1,\varphi,r) + \frac{1}{\left(f(\varphi) + r^2\right)^{\Delta_\phi}}\,,
\end{equation}
which follows from the near-boundary expansion
\begin{equation}
    \tilde G_{AdS} = \frac{z^{\Delta_\phi}}{\left(f(\varphi) + r^2\right)^{\Delta_\phi}} + \dots\ .
\end{equation}

Having summarized the method — which was fully developed in \cite{Parisini:2023nbd} — we now proceed to discuss its results. After extracting the two-point function $g_\beta(\tau)$ from the numerical solution via \eqref{H-gbeta-relation}, we fit its small-$\tau$ behavior to the OPE expansion, with coefficients kept unknown. The fitting is performed on the interval $[\tau_{\text{min}},\tau_{\text{max}}]$ for various values of $\tau_{\text{min}}$ and $\tau_{\text{max}}$. Each choice of the interval gives a fit for coefficients $a_\mathcal{O}$. We collect these different fits and take their minimum and maximum, $a_\mathcal{O}^{\text{min}}$ and $a_\mathcal{O}^{\text{max}}$. For the grid of size $N=140$, the results read as follows
\begin{equation}\label{estimates-wave-numerics}
    a_{\phi^2} \in [1.09,1.10]\,, \qquad a_{[\phi\phi]_{2\Delta_\phi+2}} \in [8.84,10.97]\ .
\end{equation}
In the fitting procedure, we have inputted exact values of the stress-tensor and double-stress-tensor one-point coefficients
\begin{equation}\label{first-two-stress-tensors}
    \Lambda_1 = \frac{3\pi^4}{80} = 3.6528...\,, \qquad \Lambda_2 = \frac{479\pi^8}{268800}\ .
\end{equation}
In order to be consistent with \cite{Parisini:2023nbd}, we have put $\beta=\pi$ in the last expressions, rather than $\beta=1$ used in the main text. The lowest double-trace coefficient in \eqref{estimates-wave-numerics} is consistent with the value found in \cite{Parisini:2023nbd}. Furthermore, had we not used the exact value for $\Lambda_1$ from \eqref{first-two-stress-tensors}, but instead treated it as an unknown in the fitting, we would have obtained
\begin{equation}\label{Lambda-1-fit}
    \Lambda_1^{\text{fit}} \in [3.30,3.77]\,,
\end{equation}
consistent with the actual value. These facts lend support to the numerical procedure employed. Finally, the asymptotic model value of the second double trace coefficient is $a^{\text{AM}}_{[\phi\phi]_{2\Delta_\phi+2}}\approx7.55$. Both the last number and the numerical result \eqref{estimates-wave-numerics} can only serve as rough guides for the true value of the holographic coefficient  $a_{[\phi\phi]_{2\Delta_\phi+2}}$. In the case of the former, this comes from the fact that asymptotic rather than exact stress-tensor data is used for the computation of $a^{\text{AM}}_{[\phi\phi]_{2\Delta_\phi+2}}$. As for the latter, the numerical procedure from above needs improvements in order to give a reliable prediction for the double stress-tensor coefficient. Nevertheless, we find that the comparison of the two numbers is promising and invites for improved estimates from both methods.

\smallskip

\bibliographystyle{JHEP}
\bibliography{bibliography} 

\end{document}